\documentclass[article,shortnames]{jss}
\usepackage{natbib}
\usepackage{amsfonts}
\usepackage{amsthm, amsmath, amssymb}
\usepackage{mathrsfs}
\usepackage{graphicx, tabularx, float}
\newcommand{\indep}{\rotatebox[origin=c]{90}{$\models$}}
\usepackage{enumitem}
\usepackage{caption}
\usepackage{booktabs, multirow}
\author{Chi Zhang\\University of California, Berkeley \And 
            Jennifer Ahern\\University of California, Berkeley \AND
            Oleg Sofrygin\\University of California, Berkeley \And 
            Mark J. van der Laan\\University of California, Berkeley}
\title{tmleCommunity: A \proglang{R} Package Implementing Target Maximum Likelihood Estimation for Community-level Data}

\Plainauthor{Chi Zhang, Jennifer Ahern, Mark J. van der Laan, Oleg Sofrygin} 
\Plaintitle{tmleCommunity: A R Package Implementing Target Maximum Likelihood Estimation for Community-level Data}
\Shorttitle{\pkg{tmleCommunity}: Target Maximum Likelihood Estimation for Community-level Data} 

\Abstract{
  \label{Over the past years, many applications aim to assess the causal effect of treatments assigned at the community level, while data are still collected at the individual level among individuals of the community.  In many cases, one wants to evaluate the effect of a stochastic intervention on the community, where all communities in the target population receive probabilistically assigned treatments based on a known specified mechanism (e.g., implementing a community-level intervention policy that target stochastic changes in the behavior of a target population of communities).}
  The \pkg{tmleCommunity} package is recently developed to implement targeted minimum loss-based estimation (TMLE) of the effect of community-level intervention(s) at a single time point on an individual-based outcome of interest, including the average causal effect. Implementations of the inverse-probability-of-treatment-weighting (IPTW) and the G-computation formula (GCOMP) are also available. The package supports multivariate arbitrary (i.e., static, dynamic or stochastic) interventions with a binary or continuous outcome. Besides, it allows user-specified data-adaptive machine learning algorithms through \pkg{SuperLearner}, \pkg{sl3} and \pkg{h2oEnsemble} packages. The usage of the \pkg{tmleCommunity} package, along with a few examples, will be described in this paper.
}
\Plainkeywords{TMLE, causal inference, community-level intervention, stochastic intervention, deterministic intervention, machine learning, additive treatment effect.}

\Address{
  Chi Zhang\\
  Division of Biostatistics\\
  University of California, Berkeley\\
  Berkeley, CA 94720\\
  E-mail: \email{chi.zhang@berkeley.edu}
}

\begin{document}	


\newpage	
\section[Introduction]{Introduction}\label{Introduction}

\subsection[Motivation]{Motivation}\label{Motivation}
The literature in fields such as epidemiology, econometrics and social science on the causal impact of community-level intervention, 
is rapidly evolving, both in observational studies and randomized trials. In observation settings, there is a rich literature on assessment of causal effects of families, schools and neighborhoods on child and adolescent development \citep{brooks_aber_duncan_1997, raudenbush_willms_1995}. For instance, the problem addressed by \citet{boyd_wooden_munro_liu_have_2008} is to estimate the impact of community violence exposure on anxiety among children of African American mothers with depression. 
Similarly, randomized community trials have increased in recent years. As pointed out by \citet{oakes_2004} and \citet{steele_2016}, scientifically speaking, community randomized controlled trials (CRCT) would be a superior strategy estimate the effects of community-level exposures due to self-selection and other difficulties. One example is the MTO study, which estimates the lower-poverty neighborhood effects on crime for female and male youth \citep*{kling_ludwig_katz_2005}. Another CRCT example is the ongoing SEARCH study, which estimates the community level interventions for the elimination of HIV in rural communities in East Africa \citep{SEARCH_2013}. Despite recent statistical advances, many of the current applications still rely on estimation techniques such as random effect models (or mixed models) \citep{laird_ware_1982} and generalized estimating equations (GEE) approach \citep{liang_zeger_1986, gardiner_luo_roman_2009}. However, those methods define the causal effect of interest as a coefficient in a most likely misspecified regression model, often resulting in bias and invalid statistical inference in observational settings, and loss of efficiency in randomized community trials.

Deterministic interventions, in which each unit's treatment is set to a fixed value or a value defined by a deterministic function of the covariates, are the main strategy implemented in the current literature for the estimation of causal effects from observational data. One causal assumption needed for parameter identifiability is the positivity assumption. For example, the strong positivity assumption requires that all individuals in the population have a nonzero probability of receiving all levels of the treatment. As argued by \citet{petersen_porter_gruber_wang_van_der_laan_2010}, this strong assumption could be quite unrealistic in many cases. For example, patients with certain characteristics may never receive a particular treatment. On the other hand, a stochastic intervention is one in which each subject receives a probabilistically assigned treatment based on a known specified mechanism. Because the form of the positivity assumption needed for identifiability is model and parameter-specific, stochastic intervention causal parameters are natural candidates if requiring a weaker version of positivity compared to other causal parameters for continuous exposures. Furthermore, a policy intervention will lead to stochastic rather than deterministic interventions if the exposure of interest can only be manipulated indirectly, such as when studying the benefits of vigorous physical activity on a health outcome of interest in the elderly \citep{bembom_van_der_laan_2007}. Because it is unrealistic to enforce every elderly person to have a certain level of physical activity depending on a deterministic rule. To deal with the previous considerations, stochastic interventions could be a more flexible strategy of defining a question of interest and being better supported by the data than deterministic interventions. Thus, using stochastic intervention causal parameters is a good way of estimating causal effects of realistic policies, which could also be naturally used to define and estimate causal effects of continuous treatments or categorical multilevel treatments \citep{munoz_van_der_laan_2011}. 

As a double-robust and asymptotically efficient substitution estimator that respects global constraints of the statistical model, targeted maximum likelihood (or minimum loss-based) estimation (TMLE) provides asymptotically valid statistical inference, with potential reduction in bias and gain in efficiency \citep{van_der_laan_rubin_2006, van_der_laan_rose_2011}. In fact, there are two \proglang{R} \citep{R_Core_Team} packages that have been instructive for the development of our package: The \pkg{tmle} \citep{tmle_R} package performs parameter estimations for a single time point binary intervention for independent and identically distributed (IID) data, including the average treatment effect (ATE), controlled direct effects (CDE), and the parameters of a marginal structural model (MSM). Besides, \citet{tmlenet_R} developed another \proglang{R} package called \pkg{tmlenet}, which provides three estimators for average causal effects (and ATE) for single time point arbitrary interventions (univariate or multivariate; static, dynamic or stochastic) in the context of network-dependent (non-IID) data, including TMLE, the inverse-probability-of-treatment-weighting (IPTW) and the parametric G-computation formula (GCOMP). This package performs logistic regression through \code{glm} and \code{speedglm}.

The development of the \pkg{tmleCommunity} package for \proglang{R} was motivated by the increasing demand of a user-friendly tool to estimate the impact of community-level arbitrary exposures in community-independent data structures with a semi-parametric efficient estimator. 
The \pkg{tmleCommunity} package also extends some of the capabilities of \pkg{tmlenet} by optionally allowing flexible data-adaptive estimations through \pkg{SuperLearner}, \pkg{sl3} and \pkg{h2oEnsemble} packages, or even user-supplied machine learning algorithms. Besides, it allows for panel data transformation, such as with random effects and fixed effects. \pkg{tmleCommunity} is available on github at \url{https://github.com/chizhangucb/tmleCommunity}.

\subsection[Structure of the article]{Structure of the article}
The article focuses on the practical usage of the \pkg{tmleCommunity} through multiple examples, therefore we omit many of the technical details. For a description of the TMLE framework for independent community data with static community-level interventions, we refer to \citep{Balzer_2017}. For a description of the TMLE of the mean outcome under a stochastic shift intervention for i.i.d data, we refer to \citep{munoz_van_der_laan_2011}. In section 2 we specify the causal model through a non-parametric structural equation model (NPSEM), allowing us to define the community-level causal effect of interest for arbitrary community-level stochastic interventions as a parameter of the NPSEM, define the corresponding observed data structure, and establish the identifiability of the causal parameter from the observed data generating distribution. We allow for general types of single time-point interventions, including static, dynamic and stochastic interventions. In other words, there are no further restrictions on the intervention distributions, which could be either degenerate (for deterministic interventions) or non-degenerate (for stochastic interventions). Next, section 2 introduces two different TMLEs of the counterfactual mean outcome across communities under a community level intervention that are based on community-level  and individual-level analysis, respectively. Both TMLEs can make use of individual level data in the hierarchical setting. The first community-level TMLE is developed under a general hierarchical causal model and can incorporate some working models about the dependence structure in a community. In other words, the Super Learner library of candidate estimators for the outcome regression can be expanded to include pooled individual-level regressions based on the working model. The first TMLE also includes the case of observing one individual per community unit as a special case. The second individual-level TMLE is developed under a more restricted hierarchical model in which the additional assumption of dependence holds. 

Section 3 shows how the \pkg{tmleCommunity} package is used to estimate those parameters proposed in the prior section through a few examples, and summarizes the common features of the functions that may be useful to \pkg{tmleCommunity} users. Then section 4 uses three simulation studies to demonstrate implementation in different observational settings. Section 5 discusses the possible extensions to the methodology and the package in the future. In section 6 we answer some frequently asked questions regarding the package. 

\section[Single time-point multivariate intervention]{Single time-point multivariate intervention}\label{Single time-point multivariate intervention}

\subsection[Causal model for community level interventions]{Causal model for community level interventions}\label{Causal model for community level interventions}
Throughout this manuscript, we use the bold font capital letters to denote random vectors and matrices. In studies of community-level interventions, we begin with a simple scenario that involves randomly selecting J independent communities from some target population of communities, sampling individuals from those chosen communities, and measuring baseline covariates and outcomes on each sampled individual at a single time point. Also, the number of chosen individuals within each community is not fixed, so communities are indexed with $j = {1,2, ..., J}$ and individual within the $j^{th}$ community are indexed with $i = {1, ..., N_j}$. 

After selection of the communities and individuals, pre-intervention covariates and a post-intervention outcome are measured on each sampled unit.  Because only some of the pre-intervention covariates have clear individual-level counterpart, the pre-intervention covariates separates into two sets: firstly, let denote $W_{j,i}$ the ($1\times{p}$) vector of $p$ such individual-level baseline characteristics, and so ${\bf W}_j = (W_{j,i}: i=1,...,N_j)$ is an $(N_j{\times}p)$ matrix of individual-level characteristics; secondly let $E_j$ represent the vector of community-level (environmental) baseline characteristics that have no individual-level counterpart and are shared by all community members, including the number of individuals selected within the community (i.e., $N_j \in E_j$). 
Last, $A_j$ is the exposure level assigned or naturally occurred in community $j$ and ${\bf Y}_j = (Y_{j,i} : i=1,...,N_j)$ is the vector of individual outcomes of interest.

In order to translate the scientific question of interest into a formal causal quantity, we first specify a NPSEM with endogenous variables $X = (E, {\bf W}, A, {\bf Y})$ that encodes our knowledge about the causal relationships among those variables and could be applied in both observational setting and randomized trials \citep{pearl_1995, pearl_2009}. 
\begin{align}\label{SCMcohort_I}
U &= (U_E, U_{\bf W}, U_A, U_{\bf Y}) \sim P_U \nonumber \\ 
E &=f_E(U_E) \\ 
{\bf W} &=f_{\bf W}(E,U_{\bf W}) \nonumber \\ 
A &=f_A(E, {\bf W}, U_A) \nonumber \\ 
{\bf Y} &=f_{\bf Y}(E,{\bf W},A,U_{\bf Y}). \nonumber
\end{align}
where the $U$ components are exogenous error terms, which are unmeasured and random with an unknown distribution $P_U$.  Given an input $U$, the function $F = \{f_E, f_{\bf W}, f_A, f_{\bf Y}\}$ deterministically assign a value to each of the endogenous variables. For example, model (\ref{SCMcohort_I}) assumes that each individual's outcome $Y$ is affected by its baseline community and individual-level covariates $(E, {\bf W})$ together with its community-level intervention and unobserved factors $(A, U_{\bf Y})$. First, while we might have specification of $f_A,$, the structural equations  $f_E, f_{\bf W}, f_{\bf Y}$ do not necessarily restrict the functional form of the causal relationships, which coud be nonparametric (entirely unspecific), semiparametric or parametric that incorporates domain knowledge. Second, as summarized by \citet{Balzer_2017}, structural causal model (\ref{SCMcohort_I}) covers a wide range of practical scenarios as it allows for the following types of between-individual dependencies within a community: (i) the individual-level covariates (and outcomes) among members of a community may be correlated as a consequence of shared measured and unmeasured community-level covariates $(E, U_E)$, and of possible correlations between unmeasured individual-level error terms $(U_{\bf W}, U_{\bf Y})$, and (ii) an individual's outcome $Y_{j,i}$ may influence another's outcome $Y_{j,l}$ within community $j$, and (iii) an individual's baseline covariates $W_{j,l}$ may influence another's outcome $Y_{j,i}$. Actually, we can make an assumption about the third type of between-individual dependence, and so the structural equation $f_{\bf Y}$ will be specified under this assumption. More details will be discussed in section (\ref{Individual-level TMLE}). Third, an important ingredient of this model is to assume that distinct communities are causally independent and identically distributed. The NPSEM defines a collection of distributions $(U,X)$, representing the full data model, where each distribution is determined by $F$ and $P_U$ (i.e., $P_{U,X,0}$ is the true probability distribution of $(U,X)$). We denote the model for $P_{U,X,0}$ with ${\cal M^F}$. 

\subsection[Counterfactuals and stochastic interventions]{Counterfactuals and stochastic interventions}\label{Counterfactuals and stochastic interventions}
${\cal M^F}$ allows us to define counterfactual random variables as functions of $(U,X)$, corresponding with arbitrary interventions. For example, with a static intervention on $A$, counterfactual ${\bf Y}_a$ can be defined as $f_{\bf Y}(E, {\bf W}, a, U_{\bf Y})$, replacing the structural equation $f_A$ with the constant $a$ \citep{van_der_laan_rose_2011}. Thus, ${\bf Y}_{j,a} = (Y_{j,i,a} : i=1,...,N_j)$ represents the vector of individual-level outcomes that would have been obtained in community $j$ if all individuals in that community had actually been treated according to the exposure level $a$. More generally, we can replace data generating functions for $A$ that correspond with degenerate choices of distributions for drawing $A$, given $U = u$ and $(E, {\bf W})$, by user-specified conditional distributions of $A^*$. Such non-degenerate choices of intervention distributions are often referred to as stochastic interventions. 

First, let $g^*$ denote our selection of a stochastic intervention identified by a set of multivariate conditional distributions of $A^*$, given the baseline covariates $(E, {\bf W})$. For convenience, we represent the stochastic intervention with a structural equation $A^* = f_{A^*}(E, {\bf W}, U_{A^*})$ in terms of random errors $U_{A^*}$, and so define ${\bf Y}_{g^{*}} = f_{\bf Y}(E, {\bf W}, A^*, U_{\bf Y})$. Then ${\bf Y}_{j,g^*} = (Y_{j,i,g^*} : i=1,...,N_j)$ denotes the corresponding vector of individual-level counterfactual outcome for community $j$. Second, let $Y^c$ denote a scalar representing a community-level outcome that is defined as a aggregate of the outcomes measured among individuals who are members within a community, and so $Y^c_{g^{*}}$ is the corresponding community-level counterfactual 
of interest. One typical choice of $Y_{j,g^*}^c$ is the weighted average response among the $N_j$ individuals sampled from community $j$, i.e. $Y_{j,g^*}^c \equiv \sum_{i=1}^{N_j} \alpha_{j,i}Y_{j,i,g^*}$, for some user-specified set of weights $\alpha$ for which $\sum_{i=1}^{N_j} \alpha_{j,i} = 1$. If the underlying community size $N_j$ differs, a natural choice of $\alpha_{j,i}$ is the reciprocal of the community size (i.e., $\alpha_{j,i} = 1/N_j$). 


\subsection[Target parameter on the NPSEM]{Target parameter on the NPSEM}\label{Target parameter on the NPSEM}

We focus on community-level causal effects where all communities in the target population receive the intervention $g^{*}$, then our causal parameter of interest is given by 
\[ \Psi^F(P_{U,X,0}) 
= \mathbb{E}_{U,X}[Y^c_{g^{*}}] 
= \mathbb{E}_{U,X} \Big\{ \sum\limits_{i=1}^{N} \alpha_{i}Y_{i,g^{*}} \Big\} 
\] 
To simplify the expression,  we use $\alpha_{i}  = 1/N$ in the remainder of article.

One type of stochastic interventions could be a shifted version of the current treatment mechanism $g_0$, i.e., $P_{g^*}(A = a | E, {\bf W}) = g_0(a  -  \nu(E, {\bf W}) | E, {\bf W})$ given a known shift function $\nu(E, {\bf W})$. A simple example is a constant shift of $\nu(E, {\bf W}) = 0.5$. Another more complex type could be stochastic dynamic interventions, in which the interventions can be viewed as random assignments among dynamic rules. A simple example corresponding to the previous shift function is $P_{g^*}(A = a | E, {\bf W}) = g_0(\text{max}\{a  -  0.5, \text{min}(a)\} | E, {\bf W})$, indicating that shifted exposure $A^*$ is always bounded by the minimum of the observed exposure $A$.

One might also be interested in the contrasts of the expectation of community-level outcome across the target population of communities under different interventions, i.e., 
\[ \Psi^F(P_{U,X,0}) =  \mathbb{E}_{U,X}(Y^c_{g^{*}_1}) - \mathbb{E}_{U,X}(Y^c_{g^{*}_2}) = \mathbb{E}_{U,X} \Big\{ \frac{1}{N} \sum\limits_{i=1}^{N}Y_{i,g^{*}_1} \Big\} - \mathbb{E}_{U,X} \Big\{ \frac{1}{N} \sum\limits_{i=1}^{N}Y_{i,g^{*}_2} \Big\}\]
where $g^{*}_1$ and $g^{*}_2$ are two different stochastic interventions. 

Finally, additive treatment effect is a special case of average causal effect with two static interventions $g_1^{*}(1 | e, {\bf w}) = 1$ and $g_2^{*}(0 | e, {\bf w}) = 1$ for any $e \in E, {\bf w} \in {\bf W}$, i.e., 
\[ \mathbb{E}_{U,X}(Y^c(1)) - \mathbb{E}_{U,X}(Y^c(0)) = \mathbb{E}_{U,X}[Y^c_{g^{*}_1(1 | e, {\bf w}) = 1}] - \mathbb{E}_{U,X}[Y^c_{g^{*}_2(0 | e, {\bf w}) = 1}] \]

\subsection[Link to observed data]{Link to observed data}\label{Link to observed data}
Consider the study design presented above where for a randomly selected community, the observed data consist of the measured pre-intervention covariates, the intervention assignment, the vector of individual-level outcomes. Formally, one observation on community $j$, is coded as  
\[ O_{j,i} = (E_j, W_{j,i}, A_j, Y_{j,i})\]
which follows the typical time ordering for the variables measured on the $i^{th}$ individuals within the $j^{th}$ community.

Assume the observed data consists of $J$  independent and identically distributed copies of $\mathbf{O}_{j} = (E_j, {\bf W}_{j}, A_j, {\bf Y}_{j}) \sim P_0$, where $P_0$ is an unknown underlying probability distribution in a model space ${\cal M}^I$. Here $\mathcal{M}^I = \{ P(P_{U,X}): P_{U,X} \in \mathcal{M}^F \}$ denotes the statistical model that is the set of possible distributions for the observed data $O$ and only involves modeling $g_0$ (i.e., specification of $f_A$). The true observed data distribution is thus $P_0  = P(P_{U,X,0})$.

\subsection[Identifiability]{Identifiability}\label{Identifiability}

By defining the causal quantity of interest in terms of stochastic interventions (and target causal parameter as a parameter of the distribution $P_{U,X,0}$) on the NPSEM and providing an explicit link between this model and the observed data, we lay the groundwork for addressing the identifiability through $P_0$. 

In order to express $\Psi^F(P_{U,X,0})$ as a parameter of the distribution $P_0$ of the observed data $O$, we now need to address the identifiability of $\mathbb{E}_{U,X}[Y^c_{g^{*}}]$ by adding two key assumptions on the NPSEM: the randomization assumption so called "no unmeasured confounders" assumption (Assumption 1) and the positivity assumption (Assumption 2). The identifiability assumptions will be briefly reviewed here, for details on identifiability, we refer to see \citep{robins_1986, van_der_laan_2010, van_der_laan_2014, munoz_van_der_laan_2011}.

\textbf{Assumption 1.}
\[ A \indep {\bf Y}_a | E, {\bf W} \]
where the counterfactual random variable  ${\bf Y}_a$ represents a collection of outcomes measured on the individuals from a community if its intervention is set to $A=a$ in causal model (\ref{SCMcohort_I}), replacing the structural equation $f_A$ with the constant $a$.

\textbf{Assumption 2.}
\[ \sup\limits_{a \in \mathcal{A}} \frac{g^{*}(a | E, {\bf W})}{g(a | E, {\bf W})} < \infty, 
\hspace{0.2cm} \text{almost everywhere} \]
where $g^{*}(a | E, {\bf W}) = P_{g^{*}}(A=a|E, {\bf W})$.

Informally, \textbf{Assumption 1} restricts the allowed distribution for $P_U$ to ensure that $A$ and $Y$ shares no common causes beyond any measured variables in $X = (E, {\bf W}, A, {\bf Y})$. For example, assumption 1 holds if $U_A$ is independent of $U_Y$ , given $E, {\bf W}$. Besides, this randomization assumption implies $A^* \indep {\bf Y}_a | E, {\bf W}$. Under \textbf{Assumption 1} and \textbf{2}, jointly with the consistency assumption (i.e., $A = a$ implies ${\bf Y}_a = {\bf Y}$), $P({\bf Y}_{g^{*}} = {\bf y} | A^* = a, E = e, {\bf W} = {\bf w}) = P({\bf Y}_a = {\bf y} | A^* = a,E = e, {\bf W} = {\bf w}) = P({\bf Y}_a = {\bf y} | E = e, {\bf W} = {\bf w}) = P({\bf Y} = {\bf y} | A = a, E = e, {\bf W} = {\bf w})$, so our counterfactual distribution $P({\bf Y}_{g^{*}} = {\bf y})$ can be written as:
\begin{align*}
P({\bf Y}_{g^{*}} = {\bf y}) &= \int_{e,{\bf w}} \int_{a} P({\bf Y}_{g^{*}} = {\bf y} | A^* = a, E = e, {\bf W} = {\bf w}) g^{*}(a|e,{\bf w}) d\mu(a) dP_{E, {\bf W}}(e, {\bf w})  \\
&= \int_{e,{\bf w}} \int_{a} P({\bf Y}_{a} = {\bf y} | E = e, {\bf W} = {\bf w}) g^{*}(a|e,{\bf w}) d\mu_a(a) dP_{E, {\bf W}}(e, {\bf w}) \\
& \text{by \textbf{assumption 1} and } A^* \indep {\bf Y}_a | E, {\bf W}  \\
&= \int_{e,{\bf w}} \int_{a} P({\bf Y} = {\bf y} | A=a, E=e, {\bf W}={\bf w}) g^{*}(a|e,{\bf w}) d\mu_a(a) dP_{E, {\bf W}}(e, {\bf w}) \\
& \text{by consistency assumption}
\end{align*}
with respect to some dominating measure $\mu_a(a)$. 

Then, $\mathbb{E}_{U,X}[{\bf Y}_{g^{*}}]$ is identified by the G-computational formula \citep{robins_1986}:
\begin{align*}
\mathbb{E}_{U,X}[{\bf Y}_{g^{*}}] &= \mathbb{E}_{E, {\bf W}}[ \mathbb{E}_{g^{*}}[{\bf Y} | A^* = a, E, {\bf W}] ] \\
&= \int_{e,{\bf w}} \int_{a} \mathbb{E}_{g^{*}}({\bf Y} | a,e,{\bf w})g^{*}(a|e,{\bf w}) d\mu_a(a) dP_{E, {\bf W}}(e, {\bf w}) 
\end{align*}

This provides us with a general identifiability result for $\mathbb{E}_{U,X}[Y^c_{g^{*}}]$, the causal effect of the community-level stochastic intervention on any community-level outcome $Y^c$ that is some real valued function of the individual-level outcome ${\bf Y}$:
\[ \mathbb{E}_{U,X}[Y_{g^{*}}^c] = \mathbb{E}_{U,X}[\sum\limits_{i=1}^n{\alpha_i}{Y_{g^{*},i}}] = \sum\limits_{i=1}^n{\alpha_i}\mathbb{E}_{E,{\bf W}}[\mathbb{E}_{g^{*}}[Y_i | A^*, E, {\bf W}]] \equiv \Psi^I(P_0) = \psi^I_0\]

\subsection[The statistical parameter and model for observed data]{The statistical parameter and model for observed data}\label{The statistical parameter and model for observed data}
If we only assume the randomization assumption in the previous section, then the statistical model $\mathcal{M}^I$ is nonparametric. Based on the result of identifiability, we note that $\Psi^I: \mathcal{M}^I \rightarrow \mathbb{R}$ represents a mapping from a probability distribution of $\mathbf{O}$ into a real number, and $\Psi^I(P_0)$ denotes the target estimand corresponding to the target causal quantity $\mathbb{E}_{U,X}[{\bf Y}_{g^{*}}]$.

Before defining the statistical parameter, we introduce some additional notation. First, we denote the marginal distribution of the baseline covariates $(E, {\bf W})$ by $Q_{E, {\bf W}}$, with a well-defined density $q_{E, {\bf W}}$, with respect to some dominating measure $\mu_y(y)$. There is no additional assumption of independence for $Q_{E, {\bf W}}$. Second, let $G$ denote the observed exposure conditional distribution for $A$ that has a conditional density $g(A | E,{\bf W})$.
Third, we assume that all $Y$ within a community are sampled from the distribution $Q_{\bf Y}$ with density given by $q_{\bf Y}({\bf Y} | A, E, {\bf W})$, conditional on the exposure and the baseline covariates $A, E, {\bf W}$. Now we introduce the notation $P = P_{\tilde{Q},G}$ for $\tilde{Q} = (Q_{\bf Y}, Q_{E,{\bf W}})$, and the statistical model becomes $\mathcal{M}^I = \{P_{\tilde{Q},G}: \tilde{Q} \in \tilde{\mathcal{Q}}, G \in \mathcal{G}\}$, where $\tilde{\mathcal{Q}}$ and $\mathcal{G}$ denote the parameter space for $\tilde{Q}$ and $G$, respectively, and $\tilde{\mathcal{Q}}$ here is nonparametric. 

Next, we define $G^*$ as the user-supplied intervention with a new density $g^{*}$, which will replace the observed conditional distribution $G$. So $G^*$ is a conditional distribution that describes how each intervened treatment is produced conditional on the baseline covariate $(E, {\bf W})$. Given $\tilde{Q}$ and $G^*$, we use $\mathbf{O}^* = (O^*_{j,i} = (E_j, W_{j,i}, A^*_j, Y^*_{j,i}): i=1,...,N_j, j=1,,,.,J)$ to denote a random variable generated under the post-intervention distribution $P_{\tilde{Q},G^*}$. Namely, $P_{\tilde{Q},G^*}$ is the post-intervention distribution of observed data $\mathbf{O}$ under stochastic intervention $G^*$ \citep{robins_1986},  and the likelihood for $P_{\tilde{Q},G^*}$ can be factorized as: 
\begin{align}\label{Likelihood 1}
p_{\tilde{Q},G^*}(\mathbf{O}^*) = [\prod\limits_{j=1}^{J} q_{\bf Y}({\bf Y}_j^{*} | A_j^*, {\bf W}_j, E_j)][\prod\limits_{j=1}^{J} g^{*}(A_j^* | E_j, {\bf W}_j)]q_{E,{\bf W}}(E,{\bf W})
\end{align}

Thus our target statistical quantity is now defined as  $\psi_0^I = \Psi^I(P_0) = \mathbb{E}_{\tilde{q}_0, g^*}[Y^c_{g^{*}}]$, where $\Psi^I(P_0)$ is the target estimand of the true distribution of the observed data $P_0 \in \mathcal{M}^I$ (i.e., a mapping from the statistical model $\mathcal{M}^I$ to $\mathbb{R}$). We define $\bar{Q}(A_j, E_j, {\bf W}_j) = \int_{y} q_{\bf Y}({\bf y} | A_j, {\bf W}_j, E)d\mu_y(y)$ as the conditional mean evaluated under common-in-$j$ distribution $Q_{\bf Y}$, and so $\bar{Q}^c(A, E, {\bf W}) \equiv E(Y^c | A,E,{\bf W})$ as the conditional mean of the community-level outcome. Now we can refer to $Q_0 = (\bar{Q}_0^c, Q_{E, {\bf W}, 0})$ as the part of the observed data distribution that our target parameter is a function of (i.e., with a slight abuse of notation $\Psi^I(P_0) = \Psi^I(Q_0)$), the parameter $\psi_0^I$ can be written as:
\begin{align}\label{parameter_likelihood}
\psi_0^I = \int_{e \in \mathcal{E}, {\bf w} \in \mathcal{W}} \int_{a \in \mathcal{A}} \bar{Q}_0^c(a, e, {\bf w}) g^{*}(a|e, {\bf w})d\mu_a(a) q_{E,{\bf W}, 0}(e, {\bf w})d\mu_{e, w}(e, {\bf w})
\end{align}
with respect to some dominating measures $\mu_a(a)$ and $\mu_{e,w}(e, {\bf w})$, where $(\mathcal{A}, \mathcal{E}, \mathcal{W})$ is the common support of $(A, E, {\bf W})$.

\subsection[Estimation and inference under general hierarchical causal model]{Estimation and inference under general hierarchical causal model}\label{Estimation and inference under general hierarchical causal model}
In the previous section, we have defined a statistical model $\mathcal{M}^I$ for the distribution of $\mathbf{O}$,  and a statistical target parameter mapping $\Psi^I$ for which $\Psi^I(P_{Q, G^*})$ only depends on $Q$. Now we want to estimate $\Psi^I(Q_0)$ via a target maximum likelihood estimator (TMLE) and construct an asymptotically valid confidence interval through the efficient influence curve (EIC). Furthermore, we present a novel method for the estimation of the outcome regression in which incorporates additional knowledge about the data generating mechanism that might be known by design.

\subsubsection[Community-level TMLE]{Community-level TMLE}\label{Community-level TMLE}
As a two-stage procedure, TMLE needs to estimate both the outcome regressions $\bar{Q}_0$ and treatment mechanism $g_0$. Since TMLE solves the EIC estimating equation, its estimator inherits the double robustness property of this EIC and is guaranteed to be consistent (i.e., asymptotically unbiased) if either $\bar{Q}_0$ or $g_0$ is consistently estimated. For example, in a community randomized controlled trial $g_0$ is known to be 0.5 and can be consistently estimated, thus its TMLE will always be consistent. Besides, TMLE is efficient when both are consistently estimated. In other words, when $g_0$ is consistent, a choice of the initial estimator for $\bar{Q}_0$ that is better able to approximate the true value $\bar{Q}_0$ may improve the asymptotic efficiency along with finite sample bias and variance of the TMLE \citep{van_der_laan_rubin_2006}.

The community-level TMLE first obtains an initial estimate $\hat{\bar{Q}}^c(A, E, W)$ for the conditional mean of the community-level outcome $\bar{Q}_0^c(A,E,W)$, and also an estimate $\hat{g}(A|E,W)$ of the community-level density of the conditional treatment distribution $g(A | E, W)$. The second targeting step is to create a targeted estimator $\hat{\bar{Q}}^{c*}$ of $\bar{Q}^c_0$ by updating the initial fit $\hat{\bar{Q}}^{c}(A,E,W)$ through a parametric fluctuation that exploits the information in the density for the conditional treatment distribution. We briefly review the estimation results and statistical inference here. For further discussion, please see \citep{munoz_van_der_laan_2011, tmlenet_R, Balzer_2017}.

Given $\hat{\bar{Q}}^{c}(A_j, E_j, {\bf W}_j), \hat{g}(A_j=a | E_j, {\bf W}_j)$ and $\hat{g}^{*}(A_j = a | E_j, W_j)$ for each community $j=1,..., J$, the predictions of the community-level outcome is easily obtained by
\[ \hat{\bar{Q}}^{c*}(a, E_j, {\bf W}_j) = expit\{logit(\hat{\bar{Q}}^{c}(a, E_j, {\bf W}_j) + \hat{\epsilon}\hat{H}_j(a,E_j, {\bf W}_j))\}, \forall j=1,...,J \]

where $\hat{H}_j(a,E_j, {\bf W}_j) = \frac{\hat{g}^{*}(A_j=a | E_j, {\bf W}_j)}{\hat{g}(A_j=a | E_j, {\bf W}_j)}$ displays the community-level clever covariate and the fluctuation parameter $\epsilon$ is obtained by a logistic regression of $Y^c$ on $\hat{H}$ with offset logit($\bar{Q}^c_n$) (Another way to achieve the targeting step is to use weighted regression intercept-based TMLE, where $\epsilon$ is obtained by a intercept-only weighted logistic regression of $Y^c$ with offset logit($\bar{Q}_n^c$), predicted weights logit($\bar{Q}_n^c$) and no covariates.)

Thus our targeted substitution estimator is the weighted mean of the targeted predictions across the J communities. One natural choice is the empirical mean defined as follows:
\[  \hat{\Psi}^I(P_{\hat{Q}^{*},\hat{g}^{*}}) = \frac{1}{J} \sum\limits_{j=1}^J \int_{e_j, {\bf w}_j} \int_{a} \hat{\bar{Q}}^{c*}(a, e_j, {\bf w}_j) \hat{g}^{*}(a | e_j, {\bf w}_j) d\mu_a(a) q_{E, {\bf W}}(e_j, {\bf w}_j) d\mu_{e,w}(e_j, {\bf w}_j) \]

In practice, community-level TMLE variance is asymptotically estimated as Var$(\hat{\Psi}^I(\hat{Q}^*)) \approx \frac{\hat{\sigma}_J^2}{J}$, where $\hat{\sigma}_J^2$ is the sample variance of the estimated influence curve obtained by
\[ \hat{\sigma}_J^2 = \frac{1}{J} \sum\limits_{j=1}^J \{ D^I(\hat{Q}^*, \hat{g})(O_j)\}^2 \]

where $D^I(\hat{Q}^*, \hat{g})$ is the plug-in estimator of the efficient influence curve of $\Psi^I$ at $P_0$:
\begin{align*}
D^I(P_0)(\mathbf{O}) &= \frac{g^{*}}{g_0}(A | E, {\bf W})(Y^c - \bar{Q}^c_0(A, E, {\bf W})) \\
& + \mathbb{E}_{g^{*}}[\bar{Q}_0^c(A,E, {\bf W}) | E, {\bf W}] - \Psi^I(\mathbb{P}_{Q,g^{*}})
\end{align*}
where, 
\[ \mathbb{E}_{g^{*}}[\bar{Q}_0^c(A,E, {\bf W}) | E, {\bf W}] = \int_a \bar{Q}_0^{c*}(a, E, {\bf W}) g^{*}(a | E, {\bf W}) d\mu_a(a) \]

This quantity $\hat{\sigma}^2$ can be used to calculate p values and 95\% confidence intervals for different parameters, e.g., $\hat{\Psi}^I(\mathbb{P}_{\hat{Q}^{*},\hat{g}^{*}}) \pm 1.96\frac{\hat{\sigma}_J^2}{J}$ for the ATE.

\subsubsection[Incorporating hierarchical structure for estimating outcome mechanism]{Incorporating hierarchical structure for estimating outcome mechanism}\label{Incorporating hierarchical structure for estimating outcome mechanism}
Based on the previously defined community-level TMLE for the mean of the exposure-specific counterfactual community level outcome, we can still incorporate individual level data rather than simply community wide aggregates of that data. As discussed in section (\ref{Counterfactuals and stochastic interventions}), one typical choice of the community-level counterfactuals of interest is the weighed average response among all individuals sampled from that community, i.e., $Y^c_{j,g^*} = \sum_{i=1}^{N_j}\alpha_{j,i}Y_{j,i,g^*}$. Hence, the conditional mean of the community-level outcome can be rewritten as a weighted average of the individual-level outcomes, $\bar{Q}^c_0(A, E, {\bf W}) = \mathbb{E}(Y^c | A, E, {\bf W}) = \sum_{i=1}^{N}\alpha_{i}\mathbb{E}(Y_i | A, E, {\bf W}) \equiv \sum_{i=1}^{N}\alpha_{i}\mathbb{E}(Y_i | A, E, {\bf W}, N)$  where the community-specific sample size $N$ is a random variable that is included in the community-level baseline covariates $E$. 

Without changing the underlying structural causal model (1), estimand and efficient influence curve, we may use an individual-level working model to incorporate pooled individual-level outcome regressions as candidates in the Super Learner library for initial estimation of the expected community-level outcome $\bar{Q}^c_0(A, E, {\bf W})$ given community and individual level covariates, along with community-level exposures. Specially, we propose a working model that assumes that
\begin{align}\label{no_covariate_interference_I}
	\mathbb{E}_0(Y_i | A, E, {\bf W}) = \mathbb{E}_0(Y_i | A, E, W_i) = \bar{Q}_0(A, E, W_i)
\end{align}
for a common function $\bar{Q}_0$. In practice, this working model suggests that each individual's outcome is drawn from a common distribution that may depend on the individual's baseline covariates, together with the intervention and community-level baseline covariates presented in his or her community, but is not directly influenced by the covariates of others in the same community. 

Furthermore, the strength of the working assumptions could be weakened by encoding the knowledge of the dependence relationship among individuals within communities, namely, defining E to progressively contain a larger subset of any individual-level covariates included in $\bf W$ \citep{Balzer_2017}. For weak covariate interference, the baseline individual-level covariates of other individuals who are connected with individual i could be included into $W_i$. Let $F_i$ denote the subset of individuals whose baseline individual-level covariates affect that individual's outcome $Y_i$, where $i \in F_i$. Now we have a less restricted and more general version of (\ref{no_covariate_interference_I}) as working model:
\begin{align}\label{no_covariate_interference_II}
\mathbb{E}_0(Y_i | A, E, {\bf W}) = \bar{Q}_0(A, E, (W_l: l \in F_i))
\end{align}
for a common function $\bar{Q}_0$.

The implementation of the community-level TMLE incorporating hierarchical data is similar to the previous community-level TMLE, except that the estimation of the community-level outcome could be based on a single pooled individual level regression $Y_{j,i}$ on $(E_j, A_j, W_{j,i})$ when assuming the aforementioned working model (\ref{no_covariate_interference_I}). We note that this TMLE never claims that the individual-level working model holds, instead, it uses the working model as a means to generate an initial estimator of $\bar{Q}_0^c$.

\subsubsection[Special case where one observation per community]{Special case where one observation per community}\label{Special case where one observation per community}

We will now consider a special case where each community has only one individual (i.e., $N = 1$), and so all individual-level baseline covariates can be treated as environmental factors (i.e., $(E, {\bf W}) = E$). 

\textbf{Nonparametric structural equation model}  

Consider a NPSEM with structural equations for endogenous variables $X = (E, A, Y)$,
\begin{align}\label{SCMcohort_III}
E &=f_E(U_E) \\ 
A &=f_A(E, U_A) \nonumber \\ 
Y &=f_{\bf Y}(E, A, U_Y). \nonumber
\end{align}
with endogenous unmeasured sources of random variation $U = (U_E, U_A, U_Y)$.

\textbf{Counterfactuals} 

Let $Y_a = f_Y(E, a, U_Y)$ denote the counterfactual corresponding with setting the treatment $A = a$, thus the community-level counterfactual outcome is the same as the only observation's outcome in community $j$ (i.e., $Y_{j,a}^c \equiv Y_{j,a}$).  

\textbf{Observed data} 

Now the observed data become $O = (E, A, Y)$. We observe $J$ i.i.d observations on $O$.

\textbf{Target parameter on NPSEM} 

Consider the following parameter of the distribution of $(U, X)$:
\[ \Psi^F(P_{U,X,0}) = \mathbb{E}_{U,X}[Y^c_{g^{*}}] =\mathbb{E}_{U,X} [Y_{g^{*}}] \] 

\textbf{Identifiability Result}
\begin{align}
\Psi^F(P_{U,X,0}) &= \mathbb{E}_{U,X}[Y^c_{g^{*}}] = \mathbb{E}_{E}[\mathbb{E}_{g^{*}}[Y | A^*, E]] \equiv \Psi^I(P_0) \nonumber 
\end{align}

\textbf{The statistical parameter}

Now let $Q_0 = (\bar{Q}_0^c, Q_{E, 0})$ and so the conditional mean of the community-level outcome becomes $\bar{Q}^c(A, E) \equiv E(Y^c | A,E,) = E(Y | A,E)$, with density given by  $q_Y^c(Y | A, E)$. Then $Q_{E} \equiv P(E)$ and we assume $q_{E}$ is a well-defined density for $Q_{E}$. Also, the community-level stochastic intervention is denoted as $g(A | E) \equiv P(A | E)$. Given $Q$ and $G^*$, the post-intervention distribution $P_{Q,G^*}$ generates a new set of observation $O^* = (O^*_{j} = (E_j, A_j^*, Y_j^*): j=1,...,J)$. Applying those newly modified notations produces the likelihood:
\begin{align}
p_{Q,G^*}(O^*) &= \Big[ \prod\limits_{j=1}^{J} q_Y^c(Y_j^{*} | A_j^*, E_j) \Big] \Big[\prod\limits_{j=1}^{J} g^{*}(A_j^* | E_j)\Big]q_{E}(E) \nonumber 
\end{align}

Based on the NPSEM above and the result of identifiability, we propose the following  target statistical quantity of the distribution of $O$:
\begin{align}
\Psi^I(P_0) =  \int_{e \in \mathcal{E}} \int_{a \in \mathcal{A}} \bar{Q}_0^c(a, e) g^{*}(a|e)d\mu_a(a) q_{E, 0}(e)d\mu_e(e) \nonumber
\end{align}
with respect to some dominating measures $\mu_a(a)$ and $\mu_e(e)$.

\textbf{Estimation and inference}

Similar to the setting in the regular community-level TMLE, given $\hat{\bar{Q}}^{c}(A_j, E_j), \hat{g}(A_j=a | E_j)$ and $\hat{g}^{*}(A_j = a | E_j)$ for each community $j=1,..., J$, we have $\hat{H}_j(a, E_j) = \frac{\hat{g}^{*}(A_j=a | E_j)}{\hat{g}(A_j=a | E_j)}$ and an updated fit of community-level outcome regression is given by: 
\[ \hat{\bar{Q}}^{c*}(a, E_j) = expit\{logit(\hat{\bar{Q}}^{c}(a, E_j) + \hat{\epsilon}\hat{H}_j(a, E_j))\}, \forall j=1,...,J \]

Thus the targeted substitution estimator defined as follows:
\[  \hat{\Psi}^I(\mathbb{P}_{\hat{Q}^{*},\hat{g}^{c*}}) = \frac{1}{J} \sum\limits_{j=1}^J \int_{e_j} \int_{a} \hat{\bar{Q}}^{c*}(a, e_j) \hat{g}^{c*}(a | e_j)d\mu_a(a) q_{E}(e_j) d\mu_e(e_j) \]

Under regularity conditions, the TMLE $\hat{\Psi}^I(\mathbb{P}_{\hat{Q}^{*},\hat{g}^{*}})$ is asymptotically linear with influence curve that can be conservatively replaced by 
\begin{align*}
D^I(\mathbb{P}_0)(O) &= \frac{g^{*}}{g_0}(A | E)(Y^c - \bar{Q}^c_0(A, E)) + \mathbb{E}_{g^{*}}[\bar{Q}_0^c(A,E) | E] - \Psi^I(\mathbb{P}_{Q,g^{*}})
\end{align*}
where $\mathbb{E}_{g^{*}}[\bar{Q}_0^c(A,E) | E] = \int_a \bar{Q}^{c*}(a, E) g^{*}(a | E) d\mu_a(a)$.

Finally 95\% confidence intervals can be constructed by $\hat{\Psi}^I(\mathbb{P}_{\hat{Q}^{*},\hat{g}^{*}}) \pm 1.96\frac{\hat{\sigma}_J^2}{J}$

where $\hat{\sigma}_J^2 = \frac{1}{J} \sum\limits_{j=1}^J \{ D^I(\hat{Q}^*, \hat{g})(O_j)\}^2$.

\subsection[Estimation and inference under restricted hierarchical model with no covariate interference]{Estimation and inference under restricted hierarchical model with no covariate interference}\label{Estimation and inference under restricted hierarchical model with no covariate interference}

\subsubsection[Individual-level TMLE]{Individual-level TMLE}\label{Individual-level TMLE}
What if the third type of dependence in model (\ref{SCMcohort_I}) mentioned in section \ref{Causal model for community level interventions} is weak or even doesn't exist? This is so called "no covariate interference" \citep{prague_wang_stephens_tchetgen_degruttola_2016, Balzer_2017}, which describes that each individual's outcome $Y_i$ is sampled from the same distribution only depending on the same individual's own baseline covariate $W_i$, the baseline community-level covariates $E$, together with the community-level intervention and that individual's unobserved factors $(A, U_{Y_I})$. Under this working assumption, we have $\mathbb{E}_0(Y_i | A, E, {\bf W}) = \bar{Q}_0(A, E, W_i)$. Therefore, when background knowledge about $Q_0$ is sufficient to ensure an assumption that working model (\ref{no_covariate_interference_I}) holds, this background changes both the underlying hierarchical causal model and the identifiability results, and so the statistical model, estimand, efficient influence curve, etc. 

In this section, we assume such additional knowledge is available and so consider a new hierarchical causal sub-model which restricts the dependence of individuals in a community. The NPSEM that represents the causal relationships among those endogenous variables is now given by:
\begin{align}\label{SCMcohort_II}
E &=f_E(U_E) \nonumber \\ 
{\bf W} &=f_{\bf W}(E,U_{\bf W}) \\ 
A &=f_A(E, {\bf W}, U_A) \nonumber \\ 
Y_i &=f_Y(E,W_i,A,U_{Y_i}). \nonumber \\
U_{Y_i} &\indep U_A | E, W_i \nonumber
\end{align}

Besides, we also assume that there is a common conditional distribution of $A$ given $(E,W_i)$
across all individuals, i.e., $P(A | E, W_i) \equiv g_{I}(A | E, W)$, where $g_{I}(A | E, W)$ denotes the individual-level stochastic intervention. Recall that we may be interested in $Y_{j,g^*_I}^c \equiv \sum_{i=1}^{N_j} \alpha_{j,i}Y_{j,i,g^*_I}$, with respect to some individual-level stochastic intervention $g^*_I$. Thus, by identifiability,
\begin{align}
\Psi^F(P_{U,X,0}) &= \mathbb{E}_{U,X}(Y^c_{g^*_I}) = \mathbb{E}_{U,X}(\sum_{i=1}^{N} \alpha_{i}Y_{i,g^*_I}) \nonumber \\
&= \sum_{i=1}^{N}\alpha_{i}\mathbb{E}_{Q_{E, {\bf W}, 0}}\Big\{ \mathbb{E}_{g^*_I}[\bar{Q}_0(A, E, W_i) | E, W_i] \Big\} \equiv \Psi^{II}(P_{Q,g^*_I}) \nonumber 
\end{align}
where $\Psi^{II}: \mathcal{M}^{II} \rightarrow \mathcal{R}$ is the target statistical quantity under the key assumptions of identifiability and working assumption (\ref{no_covariate_interference_I}), and $\mathcal{M}^{II}$ is a sub-model of $\mathcal{M}^I$.

Now, the corresponding EIC is given by:
\begin{align}
D^{II}(P_0)(\mathbf{O}) = \sum_{i=1}^{N}\alpha_{i}[\frac{g^*_I}{g_{0,I}}(A | E, W_i)(Y_i - \bar{Q}_0(A, E, W_i)) + \mathbb{E}_{g^*_I}[\bar{Q}_0(A, E, W_i) | E, W_i] - \Psi^{II}(P_{Q,g^*_I})]  \nonumber 
\end{align}

Also, the individual-level density of the conditional treatment distribution, adjusting for $E$ and the individual specific covariate $W_i$, is defined as 
\begin{align}
g_I(a | e, w_i) &= E_{\bf W}[g_I(a | e, {\bf W}) | W_i = w_i] = E_{\bf W}[g_I(a | e, {\bf W}_{-i}, {\bf W}_{i}) | W_i = w_i]\nonumber \\
&= \int_{{\bf w}_{-i}} g_I(a | e, {\bf w}_{-i}, w_i)P({\bf W}_{-i} = {\bf w}_{-i} | W_{i} = w_{i})d\mu({\bf w}_{-i}) \nonumber \\
&=  \int_{{\bf w}_{-i}} g_I(a | e, {\bf w}_{-i}, w_i)P({\bf W}_{-i} = {\bf w}_{-i})d\mu({\bf w}_{-i}) \nonumber
\end{align} 
with respect to some dominating measure $\mu({\bf w}_{-i})$, and ${\bf W}_{-i}$ represents an $((N-1)\times{p})$ matrix of individual-level covariates,  which includes all individuals in the community except $i$. 

Therefore, the estimate of the individual-level stochastic intervention is given by
\begin{align}
\hat{g}_I(a | e, w_i) &= \frac{1}{J}\sum_{j=1}^{J} \int_{{\bf w}_{j,-i}} \hat{g}_I(a | e_j, {\bf w}_{j,-i}, w_{j,i})P_n({\bf W}_{j,-i} = {\bf w}_{j,-i})d\mu({\bf w}_{j,-i})
\nonumber
\end{align} 

and the substitution estimator of $\Psi^{II}(P_{Q,g^*_I})$ is defined as follows:
\begin{align}
\Psi^{II}(P_{Q,g^*_I}) = \frac{1}{J}\sum_{j=1}^{J}\sum_{i=1}^{n_j}\alpha_{j,i}\int_{e_j, w_{j,i}} \int_{a_j} \hat{\bar{Q}}^{*}(a, e_j, w_{j,i}) \hat{g}^{*}_I(a | e_j, w_{j,i}) d\mu_{a}(a) q_{E, {\bf W}}(e_j, {\bf w}_j) d\mu_{e,w}(e_j, {\bf w}_j) \nonumber 
\end{align}

\section[Implementation in the tmleCommunity package]{Implementation in the tmleCommunity package}
Estimation of average causal effects for single time point arbitrary interventions in hierarchical data with the \pkg{tmleCommunity} package is performed with the main function \code{tmleCommunity()}, along with the auxiliary function, \code{tmleCom_Options()}, setting additional options that control the estimation algorithms in the package. Note that \code{tmleCom_Options()} needs to be specified before calling the main function \code{tmleCommunity()}, otherwise the default settings of all arguments to the \code{tmleCom_Options()} function will be used in the estimation procedure.

\subsection[Specification of observed data]{Specification of observed data}
The observed data set is passed to \code{tmleCommunity} through the data argument as a data frame, with the outcome column, the exposure column(s), the baseline covariates columns and possibly the community identifier column (usually numeric values, but no factor values are supported in the package). The data arguments include \code{Ynode, Anodes, WEnodes, communityID and YnodeDet}, which are all column names or indices in the data frame data that represent the outcome variable, exposure variable(s), community and individual level baseline covariates, community identifier variable, and indicators of deterministic values of outcome \code{Ynode}, respectively. Only \code{Anodes} and \code{WEnodes} must be specified. If \code{Ynode} is left unspecified, the left-side of the regression formula in argument \code{Qform} will be treated as \code{Ynode}. If \code{YnodeDet} is not NULL (its corresponding column should be logical or binary), observations received TRUE or 1 as their \code{YnodeDet} values are assumed to have constant values for their \code{Ynode}, thus not being used in the final estimation step. If \code{communityID} is not NULL (its corresponding column could be integer, numeric or character), it supports three options in argument \code{community.step}, including \code{"community_level", "individual_level"} and \code{"PerCommunity"}. Otherwise, it assumes that the data set has no hierarchical structure (thus automatically choose the option \code{"NoCommunity"}). More details will be discussed in section \ref{Specification of estimation method for hierarchical data}.

The other optional arguments related to the data set - \code{obs.wts}, \code{community.wts}, \code{fluctuation} - are the sampling weights for each observation, the sampling weights for each community, the choice of the fluctuation working model, respectively. Besides, if \code{fluctuation} is specified as \code{"logistic"} (the default), continuous outcomes $Y \in [a, b]$ will be bounded into the linear transformed outcome prior to estimating the outcome mechanism.

\subsection[Specification of estimation method for hierarchical data]{Specification of estimation method for hierarchical data}\label{Specification of estimation method for hierarchical data}

\code{communityID, community.step} and \code{pooled.Q } are the main arguments to determine the estimation methods for hierarchical data. First, in order to preserve the hierarchical data structure, the data should contain a column with one unique identifier per community and the user must provide the \code{communityID} argument as a column name or index in data. Failing to provide \code{communityID} will force the \code{community.step} argument to be automatically set to \code{"NoCommunity"} (the default) and to pool data across all communities and treat the data as non-hierarchical.

Second, If \code{community.step} is specified as \code{"community_level"}, the observed data will be aggregated to the community-level and the estimation of the corresponding statistical parameter will be analogous to non-hierarchical data structures. Note that \code{pooled.Q} is in regard to incorporate the pairing of individual-level covariates and outcomes (also known as the working model of "no covariate inference") in community-level TMLE although the working model is not assumed to hold. In other words, when \code{community.step} is set to \code{"community_level"}, if \code{pooled.Q = TRUE}, the pooled individual-level regressions will be added as candidates in the Super Learner library for initial estimation of the outcome mechanism; If pooled.Q = FALSE, both outcome and treatment mechanisms are estimated on the community-level and use no individual-level information. Moreover, \code{community.step} could be specified as \code{"individual_level"} under the assumption that the working model of "no covariate inference" holds. Third, the stratified TMLE that fits separate outcome and exposure mechanisms for each community can be implemented by setting \code{community.step} to \code{"perCommunity"}. Examples with more details will be provided in section \ref{Simulation studies with community-level interventions}.

Last but not least, the \code{community.wts} argument can be used to provide the community-level observation weights. If setting to \code{"size.community"} (the default), each community is weighted by its (standard deviation scaled) number of individuals and this specification inflates the weight for communities who are underrepresented due to a large degree of missing data. If setting to \code{"equal.community"}, all communities will be weighted as the same. The user-specified \code{community.wts} may be passed as a matrix with 2 columns, where the first and second columns contain the identifiers of communities and the corresponding non-negative weights, respectively.

\subsection[Specification of interventions]{Specification of interventions}\label{Specification of interventions}

The user-supplied interventions are specified by the \code{f_g0, f_gstar1} and \code{f_gstar2} arguments. First, an intervention regimen that encodes model knowledge about values of Anodes is specified with the \code{f_g0} argument, which is either a function or a vector (or a matrix / data frame if exposures are multivariate) that provides true treatment mechanism under observed Anodes. If \code{f_g0} is specified as a function, a large vector (or a data frame if multivariate) of Anodes will be sampled from the \code{f_g0} function. Second, an intervention regimen of interest is defined by replacing the conditional density $g_0$ with a new user-supplied density $g^*$, and is specified by the \code{f_gstar1} argument, which takes a function of counterfactual exposures. The function must include "data" as one of its argument names, and return a vector or a data frame of counterfactual exposures evaluated based on \code{Anodes, WEnodes} (and possibly \code{communityID}) passed as a named argument \code{"data"}. In addition, the interventions defined by \code{f_gstar1} can be static, dynamic or stochastic. For example, for a data set with a binary treatment, a stochastic regime will randomly assign treatment to 30\% of the observations, and another deterministically static regime will assign treatment for every observation. The corresponding \code{f_gstar1} function can be coded as
\begin{CodeChunk}
\begin{CodeInput}
R> define_f.gstar <- function(prob.val) {
R+   f.gstar <- function(data, ...) {
R+     rbinom(n = NROW(data), size = 1, prob = prob.val)
R+   }
R+   return(f.gstar)
R+ }
R> f.gstar_stoch.0.3 <- define_f.gstar(prob.val = 0.3)
R> f.gstar_determ.1 <- define_f.gstar(prob.val = 1)
\end{CodeInput}
\end{CodeChunk}

Alternatively, a deterministic regime can be specified by passing a vector (or a matrix / data frame) to the \code{f_gstar1} argument with one element per observation (and one column per treatment variable if multivariate). Moreover, \code{f_gstar1} can be set to a numeric value if that constant exposure will be assigned to all observations. Thus, the other two ways to code \code{f.gstar_determ.1} in the example above would be
\begin{CodeChunk}
\begin{CodeInput}
R> f.gstar_vector.1 <- rep(1L, NROW(data))
R> f.gstar_const.1 <- 1L
\end{CodeInput}
\end{CodeChunk}

\subsection[Specification of estimation algorithms for outcome regressions]{Specification of estimation algorithms for outcome regressions}\label{Specification of estimation algorithms for outcome regressions}
As discussed in section \ref{Community-level TMLE}, the first-stage of the TMLE procedure is to estimate the con- ditional mean outcome $\bar{Q}_0$. A good initial fit of $\bar{Q}_0$ could reduce reliance on bias reduction from the subsequent targeting step and provide a target parameter estimate with smaller variance. The following optional arguments to the \code{tmleCom_Options()} and \code{tmleCommunity()} functions give users control over the initial estimation of $\bar{Q}_0$:

Relevant arguments in \code{tmleCom_Options()}: 
\begin{enumerate}
	\item \code{Qestimator} A string specifying the default estimator for fitting $\bar{Q}_0$, including both parametric estimations (\code{"speedglm__glm" } and \code{"glm__glm"}) and data-adaptive estimations (\code{"h2o__ensemble"} and \code{"SuperLearner"}).
	
	\item \code{h2ometalearner} A string to pass to \code{h2o.ensemble}, specifying the prediction algorithm used to learn the optimal combination of the base learners.
	
	\item \code{h2olearner} A vector of prediction algorithms for training the ensemble's base models.
	
	\item \code{SL.library} A vector of prediction algorithms to pass to \code{SuperLearner}.
	
	\item \code{CVfolds} Optional number of splits in the V-fold cross-validation step for data-adaptive estimation.
\end{enumerate}

Relevant arguments in \code{tmleCommunity()}:
\begin{enumerate}
	\item \code{Qform} Regression formula for $\bar{Q}_0$, with the form of \code{Ynode ~ Anodes + WEnodes}.
	
	\item \code{Qbounds} A vector of 2 truncated levels on continuous $Y$ and the initial estimate $\bar{Q}_n$.
	
	\item \code{alpha} A value keeping $\bar{Q}_n$ bounded away from (0,1) for logistic fluctuation.
\end{enumerate}

Note that a negative Bernoulli log likelihood could be used as a valid loss function for $\bar{Q}_0$ when setting \code{fluctuation = "logistic"}, even if $Y$ is not binary. Compared to a regular linear fluctuation, a logistic fluctuation assures that all predicted means are in (0, 1) and the corresponding estimates for $\bar{Q}$ respect the global constraints of the observed data model, which reduces bias and variance in small samples \citep{Gruber_van_der_Laan_2010}. Before performing the estimation procedure, outcomes $Y$ will be bounded by \code{Qbounds}, and then will be automatically transformed into $Y^*$, continuous outcomes bounded by (0, 1) where $Y^* = \frac{Y-a}{b-a} \in [0,1]$. If \code{Qbounds} is unspecified, the default choice of the range of $Y$, widened by 10\% of its minimum and maximum values, will be used. Besides, if outcomes $Y$ were originally transformed into $Y^*$, fitting values of the targeted estimates will be automatically mapped back to the original scale. Once \code{Qbounds} finish bounding the observed outcomes, it will be set to (1 - \code{alpha}, \code{alpha}) and used to bound the predicted values for the initial outcome mechanism.

The default estimation algorithm for $\bar{Q}_0$ is set to \code{"speedglm__glm"}, which relies on the \pkg{speedglm} package \citep{speedglm_R} and uses its function \code{speedglm.wfit} to fit parametric generalized linear models (GLM) to medium-large data sets. We note that a direct call to \code{speedglm.wfit} that requires a model matrix as input is faster than the standard call to \code{speedglm}. Another regular parametric GLM fitting functoin \code{glm.fit} (the workhorse of \code{glm}) is also available for estimating  $\bar{Q}_0$ by setting the \code{Qestimator} argument to \code{"glm__glm"}. When \code{speedglm.wfit} or \code{glm.fit} is called, logistic regression will be used for the initial estimation of $\bar{Q}_0$.

However, in a nonparametric model, the probability distribution of the data are typically unknown and thus parametric models with assumptions that do not use realistic background knowledge and not respect the global constraints of the statistcal model are easily incorrectly specified. The recommended solution for it is to use more flexible machine-learning estimators that adapt the regression function to the data without overfitting the data. \pkg{tmleCommunity} relies on the \pkg{SuperLearner} and \pkg{h2oEnsemble} packages to perform data-adaptive estimation. Based on the oracle properties of V-fold cross validation that minimizes the estimated expected squared error loss function \citep*{van_der_Laan_Polley_Hubbar_2007}, the super learning chooses the best weighted convex combination of candidate estimators in the user-specified library, possibly including both machine learning algorithms and parametric models. One of its most important advantages is that its "free lunch" principle - Including a large variety of prediction algorithms in the super learning library could increase the chance of being consistently estimated, especially when the functional form of the conditional density is unknown. Note that \pkg{h2oEnsemble} is another package that provides Super learning method and is based on the \pkg{h2o} package (usually used to build models on large datasets).

\code{Qform} can be used to specify a regression formula that includes \code{Anodes} and \code{WEnodes} for $\bar{Q}_0$. The functional form of the formula is only important to parametric estimation algorithms \code{speedglm.wfit} and \code{glm.fit}. When using data-adaptive estimation algorithms provided by \pkg{SuperLearner} and \pkg{h2oEnsemble}, all variables on the right hand side of \code{Qform} will be treated as predictor variables passed to the candidate estimators, ignoring the original regression formula. If \code{Qform} is somehow left unspecified, all variables in \code{Anodes} and \code{WEnodes} will be treated as predictor covariates. Besides, the library of the candidate estimators can specify different functions of the passed predictor variables, such as \code{SL.glm.interaction} for second order interaction terms in \pkg{SuperLearner}. For more details on creating new wrapper functions for prediction algorithms in \pkg{h2oEnsemble} and \pkg{SuperLearner}, please see \citep{h2oEnsemble_R} and \citep{SuperLearner_R},  respectively.

Finally, we realize that sometimes the use of \pkg{SuperLearner} and \pkg{h2oEnsemble} may fail due to various reasons such as constant responses, so does the use of \code{speedglm}. Therefore, the \code{tmleCommunity()} function provides an insurance mechanism for guaranteeing the estimation procedure functions normally even if the chosen algorithm fails: If \code{"h2o__ensemble"} fails, it falls back on \code{"SuperLearner"}; If \code{"SuperLearner"} fails, it falls back on \code{"speedglm__glm"}; If \code{"speedglm__glm"} fails, it falls back on \code{"glm__glm"}. However, \code{tmleCommunity()} will terminate with an error if the last solution \code{glm.fit} also fails.

We demonstrate a simple application of the \code{tmleCommunity} function using user-specified parametric models and super learning library to estimate $\bar{Q}$ in the code chunk below. In this example, we have a simulated data of 1,000 i.i.d. observations with four baseline covariates (\code{W1, W2, W3} and \code{W4}), one binary exposure(\code{A}) and continuous outcome (\code{Y}). Its true ATE value is 2.80. Code to generate the example dataset is attached in the supplementary material. 

We begin with correctly specified models for $\bar{Q}$. Note that \code{tmleCommunity} will provide results of three estimators for ATE. In this section, we only care about the non-targeted substitution estimate that uses only an estimate of $\bar{Q}$, thus we can use \code{"gcomp"} to extract the corresponding results of the MLE estimator.
\begin{CodeChunk}
\begin{CodeInput}
R> tmleCom_Options(Qestimator = "speedglm__glm")
R> tmleCom_Qc_ATE <-
R+   tmleCommunity(data = indSample.bA.cY, Ynode = "Y", Anodes = "A", 
R+                 WEnodes = c("W1", "W2", "W3", "W4"), f_gstar1 = 1,
R+                 f_gstar2 = 0, Qform = "Y ~ W1 + W2 + W3 + W4 + A",
R+                 alpha = 0.995)
R> c(tmleCom_Qc_ATE$ATE$estimates["gcomp", ],
R+   tmleCom_Qc_ATE$ATE$vars["gcomp", ])
\end{CodeInput}
\begin{CodeOutput}
[1] 2.816628 0.004813
\end{CodeOutput}
\end{CodeChunk}

What if our assumption of the parametric model for $\bar{Q}$ is incorrect? The result of the misspecified outcome regression is shown next.
\begin{CodeChunk}	
\begin{CodeInput}
R> tmleCom_Options(Qestimator = "speedglm__glm")
R> tmleCom_Qm_ATE <-
R+   tmleCommunity(data = indSample.bA.cY, Ynode = "Y", Anodes = "A",
R+                 WEnodes = c("W1", "W2", "W3", "W4"), f_gstar1 = 1,
R+                 f_gstar2 = 0, Qform = "Y ~ W1 + A", alpha = 0.995)
R> c(tmleCom_Qm_ATE$ATE$estimates["gcomp", ],
R+   tmleCom_Qm_ATE$ATE$vars["gcomp", ])
\end{CodeInput}
\begin{CodeOutput}
[1] 3.45848993 0.01460869
\end{CodeOutput}
\end{CodeChunk}

Next, suppose we do not know its parametric model and need to use the super learning algorithm to estimate $\bar{Q}$. Instead of using the default library, we specify one that contains three prediction algorithms: \code{SL.glm}, \code{SL.bayesglm} and \code{SL.gam} (Note that \code{SL.gam} calls the \code{gam} function in the suggested \code{gam} package that uses generalized additive models \citep{gam_R}).
\begin{CodeChunk}	
\begin{CodeInput}
R> require("SuperLearner")
R> tmleCom_Options(Qestimator = "SuperLearner", CVfolds = 5,
R+                 SL.library = c("SL.glm", "SL.bayesglm", "SL.gam"))
R> tmleCom_QSL_ATE <-
R+   tmleCommunity(data = indSample.bA.cY, Ynode = "Y", Anodes = "A",
R+                 WEnodes = c("W1", "W2", "W3", "W4"), f_gstar1 = 1,
R+                 f_gstar2 = 0, rndseed = 12345)
R> c(tmleCom_QSL_ATE$ATE$estimates["gcomp", ],
R+   tmleCom_QSL_ATE$ATE$vars["gcomp", ])
\end{CodeInput}
\begin{CodeOutput}
[1] 2.809818 0.004797
\end{CodeOutput}
\end{CodeChunk}

\subsection[Specification of estimation algorithms for treatment mechanisms]{Specification of estimation algorithms for treatment mechanisms}\label{Specification of estimation algorithms for treatment mechanisms}
After finishing the initial fit for $\bar{Q}$ in the first stage of TMLE, the next step is to modify the initial estimator by using the estimation of nuisance parameter $g_0$, in order to make an optimal bias-variance trade off. Recall that the estimate $g_n$ will be used in a clever covariate that defines a parametric fluctuation model to update the initial estimate of $\bar{Q}$. The following arguments to the \code{tmleCom_Options()} and \code{tmleCommunity()} functions provide flexibility in how to choose the estimator for $g_0$:

Relevant arguments in \code{tmleCom_Options()}:
\begin{itemize}
	\item \code{gestimator} A string specifying the default estimator for fitting $g_0$,similar to \code{Qstimator} (except that \code{gestimator} also supports \code{"sl3_pipelines"},  another data-adaptive estimation method).
	
	\item \code{bin.method} Specify the method for choosing bins when discretizing the conditional continuous exposure variable, including \code{"equal.mass", "equal.len"} and \code{"dhist"}.
	
	\item \code{nbins} Number of bins when discretizing a continuous variable.
	
	\item \code{maxncats} The maximum number of unique categories a categorical variable can have.
	
	\item \code{maxNperBin} The maximum number of observations in each bin.
	
	\item \code{poolContinVar}  Logical, when fitting a model for binarized continuous variable, if pooling bin indicators across all bins and fit one pooled regression or not
	
	\item \code{savetime.fit.hbars} Logical, if skipping estimation and prediction of exposure mechanism or not, when \code{f.gstar1 = NULL} and \code{TMLE.targetStep = "tmle.intercept"}.
\end{itemize}

Relevant arguments in \code{tmleCommunity()}:
\begin{itemize}
	\item \code{hform.g0} Regression formula for $g_0$, with the form of \code{Anode ~ WEnodes}.
	
	\item \code{hform.gstar} Regression formula for the user-supplied intervention $g^*$, with the form of \code{Anode ~ WEnodes}.
	
	\item \code{lbound} One value for bounds on the ratio of the estimate of $g^*$ to the estimate of $g_0$.
	
	\item \code{h.g0_GenericModel} An object of \code{GenericModel} \pkg{R6} class containing the previously
	fitted models for $P(A|W, E)$ under observed treatment mechanism $g_0$.
	
	\item \code{h.gstar_GenericModel} An object of \code{GenericModel} \pkg{R6} class containing the previously
	fitted models for $P(A^*|W, E)$ under observed treatment mechanism $g^*$.
	
	\item \code{TMLE.targetStep}  TMLE targeting step method, either \code{"tmle.intercept" } (Default) or \code{"tmle.covariate"}.
\end{itemize}

The options for selecting estimation algorithms for the treatment mechanism are similar to those for estimating $\bar{Q}$, and they share the same choices of \code{h2ometalearner, h2olearner, SL.library}, and \code{CVfolds}. 
Beyond that, users can also the \pkg{sl3} package to perform data-adaptive estimation for $g_0$. Note that \pkg{sl3} is a  modern implementation of the Super Learner algorithm for ensemble learning and model stacking. For model details on creating new wrapper functions in \pkg{sl3}, please see \citep{sl3_R}. In order to provide a similar insurance mechanism for the estimation process of $g_0$, even if the chosen algorithm fails, the \code{tmlecommunity}() function will use the following rule: If \code{"h2o__ensemble"} fails, it falls back on \code{"sl3_pipelines"}; If \code{"sl3_pipelines"} fails, it falls back on \code{"SuperLearner"}; The rest of the mechanism will be the same as in the estimation process of $\bar{Q}_0$

Also, $g_0$ can be either estimated by parametric models or data-adaptive algorithms, depending on the choices of \code{gestimator}. Though the estimation algorithms for $\bar{Q}$ and $g_0$ could be different as long as the choices of \code{Qestimator} and \code{gestimator} differ, the same library is used for all factors of $g$. For example, the initial fit of $\bar{Q}$ is estimated by \pkg{h2oEnsemble} with \code{"h2o.glm.wrapper"} and \code{"h2o.randomForest.wrapper"}; A super learning library, including \code{SL.loess } (with \code{span = 0.8}), \code{SL.glmnet}, \code{SL.knn.20} (with neighborhood size \code{k = 20}) and \code{SL.step}, will be used for each factor of $g$.

It is important to choose the number and position of the bins when discretizing a continuous exposure variable, as the choices affect the variance of the density estimation \citep{scott_1992}. Fortunately, the type of each variable will be automatically detected (can be binary, categorical, or continuous) based on the user-specified \code{maxncats} argument. Recall that \code{maxncats} provides the maximum number of unique categories a categorical variable can have. So if one variable has more unique categories, it is automatically considered as a continuous variable.

According to \citet{Denby_Mallows_2009}, a histogram can be used as a graphically descriptive tool where its location of the bins is determined by cutting the empirical cumulative distribution function (ecdf) by a set of parallel lines. First, the \code{nbins} argument is a tuning parameter that determines the total number of bins of discretization. A cross-validation selector can be applied to data-adaptively select the candidate number of bins, which minimizes variance and maximizes precision. Note that we do not recommend too many bins due to easily violating the positivity assumption.

Next, given a number of bins, we need to choose the most convenient locations (cutoffs) for the bins. There are three alternative approaches that use a histogram to define the bin cutoffs for a continuous variable: equal-mass, equal-length, and a combination of these two methods. In \pkg{tmleCommunity} package, the choice of methods \code{bin.method} together with the other arguments \code{nbins} and \code{maxNperBin} can be used to define the values of bin cutoffs. Note that \code{maxNperBin} provides a user-specified maximum number of observations in each bin.

The default discretization method, equal mass (aka equal area) interval method, is set by passing an argument \code{bin.method="equal.mass"} to \code{tmleCom_Options()} prior to calling the main function \code{tmleCommunity()}. The interval are defined by spanning the support of $A$ into non-equal length of bins, each containing (approximately) the same number of observations. It is data-adaptive since it tends to be wide where the population density is small, and narrow where the density is large. If \code{nbins} is \code{NA} (or is smaller than $\frac{n}{\text{\code{maxNperBin}}}$), \code{nbins} will be (re)set to the integer value of $\frac{n}{\text{\code{maxNperBin}}}$ where $n$ is the total number of observations, and the default setting of \code{maxNperBin} is 500 observations per interval. This method could identify spikes in the density, but oversmooths in the tails and so could not discover outliers.

Besides, equal length interval method is set by passing an argument \code{bin.method="equal.len"} to \code{tmleCom_Options()}. The intervals are defined by spanning the support of a continuous variable into \code{nbins} number of equal length of bins. This method describes the tails of the density and identifies outliers well, but oversmooths in regions of high density and so is poor at identifying sharp peaks. Moreover, as an alternative to find a compromise between equal mass and equal length approaches, the combination method is set by passing an argument \code{bin.method="dhist"}, where \code{dhist} is named for diagonally cut histogram. For consistency, We choose the slope $a = 5 \times \text{IQR(A)}$ as suggested by \citet{Denby_Mallows_2009}.

Similar to \code{Qform}, formulae that include \code{WEnodes} can be specified for estimating components of $g$ and $g^*$ using the \code{hform.g0} and \code{hform.gstar} arguments. The functional form of the formulae is unimportant when the data-adaptive estimation algorithms are used. Also, if the \code{hform.g0} and \code{hform.gstar} arguments are unspecified, the formulae will default to main term regressions that includes all variables in \code{WEnodes}.

The \code{lbound} argument is a tuning parameter, conforming with the theoretical assumption 2 in section \ref{Identifiability} that the ratio of $g^*(a|E,{\bf W})$ to $g(a|E,{\bf W})$ must be bounded away from $+\infty$. Since the function $g^*(a|E,{\bf W})$ is user given, we can try to define it in a way so that it could be used to answer the causal question of interest, and yet it does not produce unstable weights. However, when there are unstable weights that cause extremely large value of $\frac{g^*(a|E,{\bf W})}{g(a|E,{\bf W})}$, this lack of identifiability will result in the estimates with high variance  \citep{van_der_laan_rubin_2006}. A common approach to reduce the variance of the consequent estimates is bounding $\frac{g^*(a|E,{\bf W})}{g(a|E,{\bf W})}$ away from the extremely large value, e.g., $0 \leq \frac{g^*(a|E,{\bf W})}{g(a|E,{\bf W})} \leq \frac{1}{\text{\code{lbound}}}$. However, truncation comes at a price of bias since the consistency of the estimator of $g(a|E,{\bf W})$ may be affected. Therefore, the \code{lbound} argument should be chosen carefully (it defaults to 0.005). 

\code{TMLE.targetStep} specifies how to use weights $\frac{h_{g^*}}{h_{g_N}}$ in the TMLE targeting step. If it is set to \code{"tmle.intercept"} (default), it performs the weighted intercept-based TMLE that runs a intercept-only weighted logistic regression using offsets logit($Q^*$) and weights $\frac{h_{g^*}}{h_{g_N}}$  and so no covariate. If setting to \code{"tmle.covariate"}, it performs the unweighted covariate-based TMLE that run an unweighted logistic regression using offsets logit($Q^*$) and a clever covariate $\frac{h_{g^*}}{h_{g_N}}$.

The following example illustrates IPTW estimation of the average causal effect of individual-based continuous intervention at a single time point. A sample of 5,000 is generated, with each row i consisting of four baseline covariates (\code{W1, W2, W3} and \code{W4}), one continuous exposure (\code{A}) and continuous outcome (\code{Y}). The true value for the marginal treatment effect of the intervention for the simulated data is $\psi_0 = 3.46601$. For details on code to generated the example dataset, please see the supplementary material.

Suppose we are interested in estimating the mean outcome $\psi_0$ under a truncated stochastic intervention $g^*$, which is defined by shifting the normal density of observed A until $\frac{g^*}{g_0} \geq 10$ and then its truncated to be equal to $g_0$. In this case, the \code{tmleCommunity()} function receives only one user-specified intervention, and we should utilize \code{\$EY_gstar1} to extract the results of estimates under the intervention \code{f_gstar1}. Moreover, we use \code{"iptw"} to display the results of the IPTW estimator since it relies only on the esimate of $g$. Let's begin with correctly specified models for $g_0$ and $g^*$ with a shift of 2 on the observed $A$, using the equal mass method that discretizes $A$ into 5 bins (all default choices).
\begin{CodeChunk}
\begin{CodeInput}
R> define_f.gstar <- function(shift.val, truncBD, rndseed = NULL) {
R+   f.gstar <- function(data, ...) {
R+     set.seed(rndseed)
R+     A.mu <- 0.86 * data$W1 + 0.93 * data$W3 * data$W4 + 0.41 * data$W4
R+     untrunc.A <- rnorm(n = nrow(data), mean = A.mu + shift.val, sd = 1)
R+     r.new.A <- exp(0.5 * shift.val * (untrunc.A - A.mu - shift.val / 2))
R+     trunc.A <- ifelse(r.new.A > truncBD, untrunc.A - shift.val, untrunc.A)
R+     return(trunc.A)
R+   }
R+   return(f.gstar)
R+ }
R> f.gstar <- define_f.gstar(shift.val = 2, truncBD = 10, rndseed = 1)
R>
R> gform.C <- "A ~ W1 + W3 * W4"  # correct gform
R> N <- NROW(indSample.cA.cY)
R> tmleCom_Options(gestimator = "speedglm__glm", bin.method = "equal.mass", 
R+                 nbins = 5, maxNperBin = N)
R> tmleCom_gc_default <-
R+   tmleCommunity(data = indSample.cA.cY, Ynode = "Y", Anodes = "A",
R+                 WEnodes = c("W1", "W2", "W3", "W4"), f_gstar1 = f.gstar,
R+                 rndseed = 1, hform.g0 = gform.C, hform.gstar = gform.C)
R> c(tmleCom_gc_default$EY_gstar1$estimates["iptw", ],
R+   tmleCom_gc_default$EY_gstar1$vars["iptw", ])
\end{CodeInput}
\begin{CodeOutput}
[1] 3.4406569 0.0120417
\end{CodeOutput}
\end{CodeChunk}

Note that if the discretization method is equal mass (i.e., \code{bin.method = "equal.mass"}), and each bin is allowed to contain no more than 250 observations (i.e., \code{maxNperBin} = 250), the number of bins ( or regressions) will be set to the larger value between the nearest interger of $\frac{n}{250}$ and the value of \code{nbins}, where $n$ is the total number of observations. Thus, even if \code{nbins} defaults to 5, the real number of bins for a sample of 5000 will be 20. It is worth mentioning that during the estimation, $-\infty$ and $+\infty$ are added as leftmost and rightmost cutoff points to make sure all future data points end up in one of the intervals. For example, if the real number of bins is 20, then the returned results will include 22 fitted models. However, the selection of \code{maxNperBin} doesn't have influence on the real number of bins when using the equal length and combination methods.
\begin{CodeChunk}	
\begin{CodeInput}
R> tmleCom_Options(maxNperBin = 250, bin.method = "equal.mass")
R> tmleCom_gmain_eqmass <-
R+   tmleCommunity(data = indSample.cA.cY, Ynode = "Y", Anodes = "A",
R+                 WEnodes = c("W1", "W2", "W3", "W4"), f_gstar1 = f.gstar)
R>  h.g0_models_mass <- tmleCom_gmain_eqmass$EY_gstar1$h.g0_GenericModel
R> length(h.g0_models_mass$getPsAsW.models()$`P(A|W).1`$bin_nms)
\end{CodeInput}
\begin{CodeOutput}
[1] 22
\end{CodeOutput}
		
\begin{CodeInput}
R> tmleCom_Options(maxNperBin = 250, bin.method = "equal.len")
R> tmleCom_gmain_eqlen <-
R+   tmleCommunity(data = indSample.cA.cY, Ynode = "Y", Anodes = "A",
R+                 WEnodes = c("W1", "W2", "W3", "W4"), f_gstar1 = f.gstar)
R>  h.h.g0_models_len <- tmleCom_gmain_eqlen$EY_gstar1$h.g0_GenericModel
R> length(h.h.g0_models_len$getPsAsW.models()$`P(A|W).1`$bin_nms)
\end{CodeInput}
\begin{CodeOutput}
[1] 7
\end{CodeOutput}
\end{CodeChunk}
		
As mentioned previously, when \code{f.gstar1} is inadvertently unspecified (i.e., \code{f.gstar1 = NULL}) and \code{TMLE.targetStep = "tmle.intercept"}, setting \code{savetime.fit.hbars} to \code{TRUE} allows the TMLE process to skip the estimation and prediction of exposure mechanism $P(A|W,E)$ under $g_0$ and $g^*$. It will directly set $\frac{h_{g^*}}{h_{g_N}}$ to 1 for all observations. 
\begin{CodeChunk}
\begin{CodeInput}
R> tmleCom_nofgstar <-
R>   tmleCommunity(data = indSample.cA.cY, Ynode = "Y", Anodes = "A",
R+                 WEnodes = c("W1", "W2", "W3", "W4"), f_gstar1 = NULL,
R+                 Qform = "Y ~ W1 + W2 + W3 + W4 + A")
R> c(tmleCom_nofgstar$EY_gstar1$h.g0_GenericModel,
R+   tmleCom_nofgstar$EY_gstar1$h.gstar_GenericModel)
\end{CodeInput}
\begin{CodeOutput}
[1] NULL
\end{CodeOutput}
\end{CodeChunk}

If instead we would like to estimate the same parameter except using machine learning algorithms. R code that uses \pkg{SuperLearner} to estimate $\psi_0$ is shown next. It displays a satisfactory result of estimation with a super learning library containing \code{"SL.glm", "SL.gam" } and \code{"SL.randomForest"}.
\begin{CodeChunk}
\begin{CodeInput}
R> require("SuperLearner")
R> tmleCom_Options(gestimator = "SuperLearner", maxNperBin = N,
R+                 SL.library = c("SL.glm", "SL.gam"))
R> tmleCom_gSL_default <-
R+   tmleCommunity(data = indSample.cA.cY, Ynode = "Y", Anodes = "A",
R+                 WEnodes = c("W1", "W2", "W3", "W4"), f_gstar1 = gstar,
R+                 Qform = "Y ~ W1 + W2 + W3 + W4 + A", rndseed = 1)
R> c(tmleCom_gSL_default$EY_gstar1$estimates["iptw", ],
R+   tmleCom_gSL_default$EY_gstar1$vars["iptw", ])
\end{CodeInput}
\begin{CodeOutput}
[1] 3.4471211 0.0124127
\end{CodeOutput}
\end{CodeChunk}

Another choice for performing maching leraning algorithms, especially stacked ensemble learning, is to use the  \pkg{sl3} package. Example R code for estimating $\psi_0$ is shown next. Both \code{Lrnr_glm_fast} and \code{Lrnr_glmnet} are used in the library.
\begin{CodeChunk}
\begin{CodeInput}
R> require("sl3"); require("SuperLearner")
R> tmleCom_Options(gestimator = "sl3_pipelines", maxNperBin = N, CVfolds = 5,
R>                 sl3_learner = list(glm_fast = make_learner(Lrnr_glm_fast),
R>                                    glmnet = make_learner(Lrnr_glmnet)), 
R>                 sl3_metalearner = make_learner(
R>                   Lrnr_optim, loss_function = loss_squared_error,
R>                   learner_function = metalearner_logistic_binomial))
R> tmleCom_gsl3_default <- 
R>   tmleCommunity(data = indSample.cA.cY, Ynode = "Y", Anodes = "A", 
R>                 WEnodes = c("W1", "W2", "W3", "W4"), f_gstar1 = f.gstar, 
R>                 rndseed = 1)
R> c(tmleCom_gsl3_default$EY_gstar1$estimates["iptw", ],
R>   tmleCom_gsl3_default$EY_gstar1$vars["iptw", ])
\end{CodeInput}
\begin{CodeOutput}
[1] 3.449183 0.014578
\end{CodeOutput}
\end{CodeChunk}

Recall that the \pkg{h2oEnsemble} package could also perform Super learning methods. In this case, we apply generalized linear models with penalized maximum likelihood for both base learners that are used to train the base models for the ensemble, and the metalearner that is used to learn the optimal combination of the base learners. Specifically, the base learners will include three regressions: Lasso ($\alpha = 1$), Ridge ($\alpha = 0$) and Elastic net models with $\alpha = 0.5$.
\begin{CodeChunk}
\begin{CodeInput}
R> require("h2oEnsemble")
R> h2o.glm.1 <- function(..., alpha = 1, prior = NULL) {
R+   h2o.glm.wrapper(..., alpha = alpha, , prior=prior)
R+ }
R> h2o.glm.0.5 <- function(..., alpha = 0.5, prior = NULL) {
R+   h2o.glm.wrapper(..., alpha = alpha, , prior=prior)
R+ }
R> h2o.glm.0 <- function(..., alpha = 0, prior = NULL) {
R+   h2o.glm.wrapper(..., alpha = alpha, , prior=prior)
R+ }
R> tmleCom_Options(gestimator = "h2o__ensemble", maxNperBin = N,
R+                 h2ometalearner = "h2o.glm.wrapper",
R+                 h2olearner = c("h2o.glm.1", "h2o.glm.0.5", "h2o.glm.0"))
R> tmleCom_gh2o_default <-
R+   tmleCommunity(data = indSample.cA.cY, Ynode = "Y", Anodes = "A",
R+                 WEnodes = c("W1", "W2", "W3", "W4"),
R+                 f_gstar1 = f.gstar, rndseed = 1)
R> c(tmleCom_gh2o_default$EY_gstar1$estimates["iptw", ],
R+   tmleCom_gh2o_default$EY_gstar1$vars["iptw", ])
\end{CodeInput}
\begin{CodeOutput}
[1] 3.4321917 0.0118350
\end{CodeOutput}
\end{CodeChunk}		

\subsection[Summary of key arguments to the tmleCommunity function]{Summary of key arguments to the tmleCommunity function}\label{Summary of key arguments to the tmleCommunity function}
For full details, see the documentation for the \pkg{tmleCommuity} package (cite ***).
\begin{itemize}
	\item{\code{data}} Observed data, \code{data.frame} with named columns, containing \code{WEnodes}, \code{Anode}, \code{Ynode} and possibly \code{communityID}, \code{YnodeDet}. 
	
	\item{\code{Ynode}} Column name or index in \code{data} of outcome variable. Outcome can be either binary or continuous (could be beyond 0 and 1). If \code{Ynode} undefined, the left-side of the regression formula in argument \code{Qform} will be treated as \code{Ynode}.
	
	\item{\code{Anodes}} Column names or indices in \code{data} of exposure (treatment) variables.
	
	\item{\code{WEnodes}} Column names or indices in \code{data} of individual-level (and possibly community-level) baseline covariates. Factors are not allowed.
	
	\item{\code{YnodeDet}} Optional column name or index in \code{data} of indicators of deterministic values of outcome \code{Ynode}, coded as (\code{TRUE} / \code{FALSE}) or (1 / 0). If TRUE or 1, value of \code{Ynode} is given deterministically / constant
	
	\item{\code{obs.wts}} Optional choice to provide/ construct a vector of individual-level observation (sampling) weights (of length \code{nrow(data)}). Currently supports a non-negative numeric vector, \code{"equal.within.pop"} (Default) and \code{"equal.within.community"}. If \code{"equal.within.pop"}, weigh individuals in the entire dataset equally (weigh to be all 1); If \code{"equal.within.community"}, weigh individuals within the same community equally (i.e., 1 / (number of individuals in each community)).
	
	\item{\code{community.step}} Methods to deal with hierarchical data, user needs to specify one of the four choices: \code{"NoCommunity"} (Default), \code{"community_level"}, \code{"individual_level"}, and  \code{"PerCommunity"}. If \code{"NoCommunity"}, claim that no hierarchical structure in data; If \code{"community_level"}, use the community-level TMLE; If \code{"individual_level"}, use the individual-level TMLE cooperating with the assumption of no covariate interference; Finally if \code{"perCommunity"}, use stratified TMLE. If \code{communityID} = \code{NULL}, then automatically pool over all communities (i.e., treated it as \code{"NoCommunity"}).
	
	\item{\code{communityID}} Optional column name or index in \code{data} representing community identifier variable. If known, it can support the three options within \code{community.step}: \code{"community_level"}, \code{"individual_level"} and \code{"PerCommunity"}.
	
	\item{\code{community.wts}} Optional choice to provide/ construct a matrix of community-level observation weights (where dimension = $J{\times}2$, where $J$ = the number of communities). The first column contains the identifiers / names of communities (ie., \code{data[, communityID]}) and the second column contains the    corresponding non-negative weights. Currently only support a numeric matrix with 2 columns, \code{"size.community"} (Default) and \code{"equal.community"}. If setting \code{community.wts} = "size.community", treat the number of individuals within each community as its weight, respectively. And if \code{community.wts} = "equal.community", assumed weights to be all 1.
	
	\item{\code{pooled.Q}} Logical for incorporating hierarchical data to estimate the outcome mechanism. If \code{TRUE}, use a pooled individual-level regression for initial estimation of the mean outcome (i.e., outcome mechanism). Default to be \code{FALSE}.
	
	\item{\code{f_g0}} Optional function used to specify model knowledge about value of Anodes. It estimates $P(A | W, E)$ under \code{g0} by sampling a large vector/ data frame of \code{Anode} (of length \code{nrow(data)*n_MCsims} or number of rows if a data frame) from \code{f_g0} function.
	
	\item{\code{f_gstar1}} Either a function or a vector or a matrix/ data frame of counterfactual exposures, depending on the number of exposure variables. If a matrix/ data frame, its number of rows must be either \code{nrow(data)} or 1 (constant exposure assigned to all observations), and its number of columns must be \code{length(Anodes)}. Note that the column names should match with the names in \code{Anodes}. If a vector, it must be of length \code{nrow(data)} or 1. If a function, it must return a vector or a data frame of counterfactual exposures sampled based on \code{Anodes}, \code{WEnodes} (and possibly \code{communityID}) passed as a named argument "data". Thus, the function must include "data" as one of its argument names. The interventions defined by \code{f_gstar1} can be static, dynamic or stochastic.
	
	\item{\code{f_gstar2}} Either a function or a vector or a matrix/ data frame of counterfactual exposures, depending on the number of exposure variables. It has the same components and requirements as \code{f_gstar1} has.
	
	\item{\code{Qform}} Character vector of regression formula for Ynode. If not specified (i.e., \code{NULL}), the outcome variable is regressed on all covariates included in Anodes and WEnodes (i.e., \code{Ynode ~ Anodes + WEnodes}).
	
	\item{\code{Qbounds}} Vector of upper and lower bounds on $Y$ and predicted value for initial $Q$. Default to the range of $Y$, widened by 10\% of the min and max values.
	
	\item{\code{alpha}} Used to keep predicted values for initial $Q$ bounded away from (0,1) for logistic fluctuation (set \code{Qbounds} to (1 - \code{alpha}), \code{alpha}).
	
	\item{\code{fluctuation}} Default to "logistic", it could also be "linear" (for targeting step).
	
	\item{\code{hform.g0}} Character vector of regression formula for estimating the conditional density of $P(A | W, E)$ under observed treatment mechanism $g_0$. If not specified, its form will be \code{Anodes ~ WEnodes}. If there are more than one exposure, it fits a joint probability.
	
	\item{\code{hform.gstar}} Character vector of regression formula for estimating the conditional density $P(A | W, E)$ under user-supplied interventions \code{f_gstar1} or \code{f_gstar2}. If not specified, it use the same regression formula as used in \code{hform.g0}.
	
	\item{\code{lbound}} Value between (0,1) for truncation of predicted $P(A | W, E)$. Default to 0.005. 
	
	\item{\code{h.g0_GenericModel}} An object of \code{GenericModel} \pkg{R6} class containing the previously fitted models for $P(A | W, E)$ under observed treatment mechanism $g_0$, one of the returns of \code{tmleCommunity} function. If known, predictions for $P(A=a | W=w, E=e)$ under $g_0$ are based on the fitted models in \code{h.g0_GenericModel}.
	
	\item{\code{h.gstar_GenericModel}} An object of \code{GenericModel} \pkg{R6} class containing the previously fitted models for $P(A^* | W, E)$ under intervention \code{gstar}, one of the returns of \code{tmleCommunity} function. If known, predictions for $P(A=a | W=w, E=e)$ under \code{gstar} are based on the fitted models in \code{h.gstar_GenericModel}.
	
	\item{\code{TMLE.targetStep}} TMLE targeting step method, either \code{"tmle.intercept"} (Default) or \code{"tmle.covariate"}.
	
	\item{\code{n_MCsims}} Number of simulations for Monte-Carlo analysis. Each simulation generates new exposures under \code{f_gstar1} or \code{f_gstar2} (if specified) or \code{f_g0} (if specified), with a sample size of nrow(data). Then these generated exposures are used when fitting the conditional densities $P(A | W, E)$ and estimating for IPTW and GCOMP under intervention \code{f_gstar1} or \code{f_gstar2} . Note that deterministic intervention only needs one simulation and stochastic intervention could use more simulation times such as 10 (Default to 1).
	
	\item{\code{CI_alpha}} Significance level (alpha) used in constructing a confidence interval. Default to 0.05.
	
	\item{\code{rndseed}} Random seed for controlling sampling A under \code{f_gstar1} or \code{f_gstar2} (for reproducibility of Monte-Carlo simulations)
	
	\item{\code{verbose}} Flag. If \code{TRUE}, print status messages. Default to \code{FALSE}. It can be turned on by setting \code{options(tmleCommunity.verbose = TRUE)}.
\end{itemize}

\subsection[Summary of key arguments to the tmleComOptions function]{Summary of key arguments to the tmleComOptions function}\label{Summary of key arguments to the tmleComOptions function}
For full details, see the documentation for the \pkg{tmleCommuity} package (cite ***).
\begin{itemize}
	\item{\code{Qestimator}} A string specifying default estimator for outcome mechanism model fitting.  The default estimator is \code{"speedglm__glm"}, which estimates regressions with \code{speedglm.wfit}; Estimator \code{"glm__glm"} uses \code{glm.fit}; Estimator \code{"h2o__ensemble"} implements the super learner ensemble (stacking) algorithm using the H2O R interface; Estimator \code{"SuperLearner"} implements the super learner prediction methods. Note that if \code{"h2o__ensemble"} fails, it falls back on \code{"SuperLearner"}. If \code{"SuperLearner"} fails, it falls back on \code{"speedglm__glm"}. If \code{"speedglm__glm"} fails, it falls back on \code{"glm__glm"}.
	
	\item{\code{gestimator}} A string specifying default estimator for exposure mechanism fitting. It has the same options as \code{Qestimator}.
	
	\item{\code{bin.method}} Specify the method for choosing bins when discretizing the conditional continuous exposure variable \code{A}. The default method is \code{"equal.mass"}, which provides a data-adaptive selection of the bins based on equal mass/ area, i.e., each bin will contain approximately the same number of observations as others. Method \code{"equal.len"} partitions the range of \code{A} into equal length \code{nbins} intervals. Method \code{"dhist"} uses a combination of the above two approaches. Please see Denby and Mallows "Variations on the Histogram" (2009) for more details.
	
	\item{\code{nbins}} When \code{bin.method = "equal.len"}, set to the user-supplied number of bins when discretizing a continuous variable. If not specified, then default to 5; If setting to as \code{NA}, then set to the nearest integer of \code{nobs/ maxNperBin}, where \code{nobs} is the total number of observations in the input data. When method is \code{"equal.mass"}, \code{nbins} will be set as the maximum of the default \code{nbins} and the nearest integer of \code{nobs/ maxNperBin}.
	
	\item{\code{maxncats}} Integer that specifies the maximum number of unique categories a categorical variable \code{A[j]} can have. If \code{A[j]} has more unique categories, it is automatically considered a continuous variable. Default to 10.
	
	\item{\code{maxNperBin}} Integer that specifies the maximum number of observations in each bin when discretizing a continuous variable \code{A[j]} (applies directly when \code{bin.method =} \code{"equal.mass"} and indirectly when \code{bin.method = "equal.len"}, but \code{nbins = NA}).
	
	\item{\code{parfit}} Logical. If \code{TRUE}, perform parallel regression fits and predictions for discretized continuous variables by functions \code{foreach} and \code{dopar} in \code{foreach} package. Default to \code{FALSE}. Note that it requires registering a parallel backend prior to running \code{tmleCommunity} function, e.g., using \code{doParallel} R package and running \code{registerDoParallel(cores = ncores)} for \code{ncores} parallel jobs.
	
	\item{\code{poolContinVar}} Logical. If \code{TRUE}, when fitting a model for binarized continuous variable, pool bin indicators across all bins and fit one pooled regression. Default to \code{FALSE}.
	
	\item{\code{savetime.fit.hbars}} Logical. If \code{TRUE}, skip estimation and prediction of exposure mechanism P(A|W,E) under $g_0 \& g^*$ when \code{f.gstar1 = NULL} and \code{TMLE.targetStep = "tmle.intercept"}, and then directly set \code{h_gstar_h_gN = 1} for each observation. Default to \code{TRUE}.
	
	\item{\code{h2ometalearner}} A string to pass to \code{h2o.ensemble}, specifying the prediction algorithm used to learn the optimal  combination of the base learners. Supports both h2o and SuperLearner wrapper functions. Default to "h2o.glm.wrapper".
	
	\item{\code{h2olearner}} A string or character vector to pass to \code{h2o.ensemble}, naming the prediction algorithm(s) used to train the base models for the ensemble. The functions must have the same format as the h2o wrapper functions. Default to "h2o.glm.wrapper".
	
	\item{\code{CVfolds}} Set the number of splits for the V-fold cross-validation step to pass to \code{SuperLearner} and \code{h2o.ensemble}. Default to 5.
	
	\item{\code{SL.library}} A string or character vector of prediction algorithms to pass to \code{SuperLearner}. Default to \code{c("SL.glm", "SL.step", "SL.glm.interaction")}. For more available algorithms see \code{SuperLearner::listWrappers()}.
\end{itemize}

\section[Simulation studies with community-level interventions]{Simulation studies with community-level interventions}\label{Simulation studies with community-level interventions}

\subsection[Simulation 1 - Stochastic interventions]{Simulation 1 - Stochastic interventions}\label{Simulation 1 - Stochastic interventions}
We perform a simulation study evaluating the finite sample bias and variance of the TMLE presented in section \ref{Community-level TMLE} and \ref{Individual-level TMLE}, including both community-level and individual-level TMLE. Besides, we compare the performance of TMLE estimator with that of Inverse-Probability-of-Treatment-Weighted estimator (IPTW) and parametric G-computation formula estimator (GCOMP). In order to estimate the average causal effect of community-level intervention(s) at a single time point on an individual-based outcome, we simulate a data set consisting of 1000 independent communities, each (community $j$) containing $n_j$ (non-fixed) number of individuals where $n_j$ is drawn from a normal distribution with mean 50 and standard deviation 10 and rounded to the nearest integer. First, we sample $n_j$ i.i.d. community-level baseline covariates $(E_1, E_2)$, distributed as
\[n_j \sim N(50, 10) \hspace{1cm} E_{1,j} \sim Unif(0,1) \hspace{1cm} E_{2,j} \sim Unif\{0.2, 0.4, 0.6, 0.8\}\]

Then 3 dependent individual-level baseline covariates ($W_1,W_2,W_3$) are drawn as a function
of community-level baseline covariates, respectively.
\begin{align}
&{\bf W}_{1,n_j} \sim (Bern(expit(-0.4 + 1.2E_{1,j} - 1.3E_{2,j})))_{i=1,...,n_j} \nonumber \\
& \begin{pmatrix} {\bf W}_{2,n_j} \\ {\bf W}_{3,n_j}  \end{pmatrix} \sim 
N(\begin{array}{c} 1- 0.8E_{1,j} - 0.4E_{2,j} \\ 0.5 + 0.2E_{1,j}  \end{array}, 
\Sigma = \begin{bmatrix} 1 & 0.6 \\ 0.6 & 1 \end{bmatrix}) \nonumber
\end{align}

And a community-level continuous treatment $A$ is sampled conditionally on the values of all baseline covariates.
\begin{gather*}
A_j \sim N(-1.2 + 0.8E_1 + 0.21E_2 + 3W_{1,n_j}^c - 0.7W_{2,n_j}^c + 0.3W_{3,n_j}^c, 1) \nonumber \\
\text{where}  \hspace{0.5cm} 
W_{1,n_j}^c = \frac{1}{n_j}\sum\limits_{i=1}^{n_j}{\bf W}_{1,n_j} \hspace{1cm} 
W_{2,n_j}^c = \frac{1}{n_j}\sum\limits_{i=1}^{n_j}{\bf W}_{2,n_j} \hspace{1cm} 
W_{3,n_j}^c = \frac{1}{n_j}\sum\limits_{i=1}^{n_j}{\bf W}_{3,n_j} \hspace{1cm} 
\nonumber
\end{gather*}

Also a truncated stochastic intervention $g^*$ is defined by shifting the normal density of observed $A$ by some known constant $shift>0$ until $\frac{g^*}{g_0}$ exceeds a known constant $bound$, and otherwise, the intervention keeps the observed exposure $A$ unchanged. So the intervened exposure $A_j^*$ is distributed as 
\begin{gather*}
A_j^* = 
\begin{cases} A_j  + shift & exp\{1.5 * shift * (A_j - \mu(E_j, W_{n_j}^c)) - \frac{shift}{4}\} > truncbd \\ 
A_j, & o.w. \end{cases} \nonumber \\
\text{where}  \hspace{0.5cm} 
\mu(E_j, W_{n_j}^c) = -1.2 + 0.8E_1 + 0.21E_2 + 3W_{1,n_j}^c - 0.7W_{2,n_j}^c + 0.3W_{3,n_j}^c
\nonumber
\end{gather*}

Last, the individual-level binary outcome $Y$ that is a function of treatment and all baseline covariates is simulated. Similarly, the post-intervened outcome $Y^*$, under stochastic intervention $g^*$, is defined as
\begin{itemize}
	\item \textbf{Case 1}: Working model holds 
	\begin{gather*}
	Y_j \sim Bern(expit(-1.7 + 1.7A_j + 0.5E_{1,j} - 1.2E_{2,j} + 1.1{\bf W}_{1,n_j} + 1.3 {\bf W}_{2,n_j} - 0.4{\bf W}_{3,n_j})) \\
	Y_j^* \sim Bern(expit(-1.7 + 1.7A_j^* + 0.5E_{1,j} - 1.2E_{2,j} + 1.1{\bf W}_{1,n_j} + 1.3 {\bf W}_{2,n_j} - 0.4{\bf W}_{3,n_j})) 
	\end{gather*}
	
	\item \textbf{Case 2}: Working model is not a reasonable approximation
	\begin{gather*}
	Y_j \sim Bern(expit(-1.7 + 1.2A_j - 0.2E_{1,j} + 1.1E_{2,j} + 5.8{\bf W}_{1,n_j}^c - 3.1 {\bf W}_{2,n_j}^c - {\bf W}_{3,n_j}^c \\ 
	+ 0.4{\bf W}_{1,n_j} + 0.2 {\bf W}_{2,n_j} - 0.4{\bf W}_{3,n_j})) \\
	Y_j^* \sim Bern(expit(-1.7 + 1.2A_j^* - 0.2E_{1,j} + 1.1E_{2,j} + 5.8{\bf W}_{1,n_j}^c - 3.1 {\bf W}_{2,n_j}^c - {\bf W}_{3,n_j}^c + \\
	0.4{\bf W}_{1,n_j} + 0.2 {\bf W}_{2,n_j} - 0.4{\bf W}_{3,n_j}))
	\end{gather*}
\end{itemize}

The next code chunk shows how to simulate a data set according to the previous data generating distributions where the working model fails. The code also defines the stochastic intervention $g^*$ that we are interested in. Assuming that we want to evaluate the effect of a constant shift of 1, given a truncation bound of 5, the true parameter value under this stochastic intervention is $\psi_0 = 0.558$.

\begin{CodeChunk}
\begin{CodeInput}
R> getY <- function(A, E1, E2, W1, W2, W3, bs, n.ind) {
R+   prob.Y <- plogis(bs[1] + bs[2] * A + bs[3] * E1 + bs[4] * E2
R+                    + bs[5] * mean(W1) + bs[6] * mean(W2) + bs[7] * mean(W3)
R+                    + bs[8] * W1 + bs[9] * W2 + bs[10] * W3)
R+   rbinom(n = n.ind, size = 1, prob = prob.Y)
R+ }
R> 
R> get.cluster.Acont <- function(id, n.ind, truncBD = 5, shift = 1,
R+                               working.model = T) {
R+   # Construct community- & individual-level baseline covariates E, W 
R+   E1 <- runif(n = 1, min = 0, max = 1)
R+   E2 <- sample(x = c(0.2, 0.4, 0.6, 0.8), size = 1)
R+   prob.W1 <- plogis(- 0.4 + 1.2 * E1 - 1.3 * E2)
R+   W1 <- rbinom(n = n.ind, size = 1, prob = prob.W1)
R+   W2_mean <- 1 - 0.8 * E1 - 0.4 * E2
R+   W3_mean <- 0.5 + 0.2 * E1
R+   W2W3 <- MASS::mvrnorm(n = n.ind, mu = c(W2_mean, W3_mean),
R+                         Sigma = matrix(c(1, 0.6, 0.6, 1), ncol = 2)) 
R+   W2 <- W2W3[, 1]
R+   W3 <- W2W3[, 2]
R+   A.mu <- -1.2 + 0.8 * E1 + 0.21 * E2 + 3 * mean(W1) - 
R+     0.7 * mean(W2) + 0.3 * mean(W3)
R+   A <- rnorm(n = 1, mean = A.mu, sd = 1)
R+   untrunc.A.gstar <- A + shift
R+   r.new.A <- exp(1.5 * shift * (untrunc.A.gstar - A.mu - shift / 4)) 
R+   trunc.A.gstar <- ifelse(r.new.A > truncBD, A, untrunc.A.gstar)
R+   if (working.model) { # when working.model holds
R+     betas <- c(-1.7, 1.7, 0.5, -1.2, 0, 0, 0, 1.1, 1.3, -0.4)
R+   } else { # when working.model fails
R+     betas <- c(-1.7, 1.2, -0.2, 1.1, 5.8, -3.1, -1, 0.4, 0.2, -0.4)
R+   }
R+   Y <- getY(A, E1, E2, W1, W2, W3, betas, n.ind)
R+   Y.gstar <- getY(trunc.A.gstar, E1, E2, W1, W2, W3, betas, n.ind)
R+   return(data.frame(cbind(id, E1, E2, W1, W2, W3, A, Y, Y.gstar)))
R+ }
R>
R> get.fullDat.Acont <- function(J, n.ind, truncBD = 5, shift = 1,
R+                               working.model = T, n.fix = F, only.Y = F) {
R+   if (n.fix) {
R+     n.ind <- rep(n.ind, J)
R+   } else {  # don't fix the number of obs in each community 
R+     n.ind <- round(rnorm(J, n.ind, 10))
R+     n.ind[n.ind <= 0] <- n.ind
R+   }
R+   if (only.Y) {id <- Y <- Y.gstar <- NULL } else { full.dat <- NULL}
R+   for(j in 1:J) {
R+     cluster.data.j <- get.cluster.Acont(working.model = working.model,
R+       id = j, n.ind = n.ind[j], truncBD = truncBD, shift = shift)
R+     if (only.Y) {
R+       id <- c(id, cluster.data.j[, "id"])
R+       Y <- c(Y, cluster.data.j[, "Y"])
R+       Y.gstar <- c(Y.gstar, cluster.data.j[, "Y.gstar"])
R+     } else {
R+       full.dat <- rbind(full.dat, cluster.data.j)
R+     }
R+     if (!only.Y) { full.dat$id <- as.integer(full.dat$id) }
R+   }
R+   ifelse(only.Y,
R+          return(data.frame(cbind(id, Y, Y.gstar))), return(full.dat))
R+ }
R> 
R> PopDat.wmF <- get.fullDat.Acont(J = 4000, n.ind = 1000, truncBD = 5,
R+                                 shift = 1, working.model = F, only.Y = T) {
R> PopDat.wmF.agg <- aggregate(PopDat.wmF, by=list(PopDat.wmF$id), mean)
R> truth.wmF <- mean(PopDat.wmF.agg$Y.gstar)
R> 
R> comSample.wmF <- get.fullDat.Acont(
R+    J = 1000, n.ind = 50, truncBD = 5, shift = 1, working.model = F)
R> comSample.wmF$Y.gstar <- NULL
R>
R> define_f.gstar <- function(shift.val, truncBD, rndseed = NULL) {
R+   f.gstar <- function(data, ...) {
R+     set.seed(rndseed)
R+     A.mu <- - 1.2 + 0.8 * data$E1 + 0.21 * data$E2 +
R+       3 * mean(data$W1) - 0.7 * mean(data$W2) + 0.3 * mean(data$W3)
R+     untrunc.A <- rnorm(n = nrow(data), mean = A.mu + shift.val, sd = 1)
R+     r.new.A <- exp(1.5 * shift.val * (untrunc.A - A.mu - shift.val / 4))
R+     trunc.A <- ifelse(r.new.A > truncBD, untrunc.A - shift.val, untrunc.A)
R+     return(trunc.A)
R+   }
R+   return(f.gstar)
R+ }
R> f.gstar <- define_f.gstar(shift.val = 1, truncBD = 5)
\end{CodeInput}
\end{CodeChunk}	
	
We first demonstrate how to use the two distinct approaches for leveraging a hierarchical data structure. Recall that the first approach treats community rather than individual as the unit of analysis and performs estimation on the aggregated data. It can also incorporate hierarchical structure for estimating outcome mechanism by adding a single pooled individual-level regression in the Super Learner library. The second approach, on the other hand, runs pooled individual-level regressions on both outcome and treatment mechanisms, because it utilizes the pairing of individual-level covariates and outcomes. Note that all approaches use the equal-mass method for choosing bins (for the continuous exposure), and \code{speedglm} as the estimators for both outcome and exposure mechanisms. Parameter estimates are obtained from 200 repetitions of the simulation.
\begin{CodeChunk}
\begin{CodeInput}
R> niterations <- 200  # Number of repetitions
R> J <- 1000
R> n <- 50
R> res.wmF.Ia <- res.wmF.Ib <- res.wmF.II <-
R+   as.data.frame(matrix(NA, nrow = nReps, ncol = 9))
R> names(res.wmF.Ia) <- names(res.wmF.Ib) <- names(res.wmF.II) <-
R+   c("TMLE.est", "IPTW.est", "Gcomp.est", "TMLE.var", "IPTW.var",
R+     "Gcomp.var", "TMLE.cover", "IPTW.cover", "Gcomp.cover")
R>
R> for (i in 1:niterations) {
R+   # Generate the full hierarchical data
R+   data <- get.fullDat.Acont(J = J, n.ind = n, truncBD = 5,
R+                             shift = 1, working.model = F)
R+   tmleCom_Options(maxNperBin = NROW(data), nbins = 5)
R+
R+   # Check if the true value falls into the confidence interval
R+   getCover <- function(CI, truth) {
R+     return(as.integer(CI[, 1] <= truth & truth <= CI[, 2]))
R+   }
R+
R+  # Community-level analysis without a pooled regression on outcome
R+   tmle_comQg <- tmleCommunity(
R+     data = data, communityID = "id", Ynode = "Y", Anodes = "A",
R+     WEnodes = c("E1", "E2", "W1", "W2", "W3"), f_gstar1 = f.gstar,
R+     community.step = "community_level", pooled.Q = FALSE,
R+     obs.wts = "equal.within.community", rndseed = 1)
R+   res.wmF.Ia[i, 1:6]<- 
R+     unlist(sapply(tmle_comQg$EY_gstar1[1:2], as.vector))
R+   res.wmF.Ia[i, 7:9] <- getCover(tmle_comQg$EY_gstar1$CIs, truth.wmF)
R+
R+   # Community-level analysis with a pooled regression on outcome
R+   tmle_cQ.pg <- tmleCommunity(
R+     data = data, communityID = "id", Ynode = "Y", Anodes = "A",
R+     WEnodes = c("E1", "E2", "W1", "W2", "W3"), f_gstar1 = f.gstar,
R+     community.step = "community_level", pooled.Q = TRUE, 
R+     obs.wts = "equal.within.community", rndseed = 1)
R+   res.wmF.Ib[i, 1:6] <-
R+     unlist(sapply(tmle_cQ.pg$EY_gstar1[1:2], as.vector))
R+   res.wmF.Ib[i, 7:9] <- getCover(tmle_cQ.pg$EY_gstar1$CIs, truth.wmF)
R+
R+   # Individual-level analysis
R+   tmle_poolQg <- tmleCommunity(
R+     data = data, communityID = "id", Ynode = "Y", Anodes = "A",
R+     WEnodes = c("E1", "E2", "W1", "W2", "W3"), f_gstar1 = f.gstar,
R+     community.step = "individual_level", rndseed = 1)
R+   res.wmF.II[i, ] <-
R+     unlist(sapply(tmle_poolQg$EY_gstar1[1:2], as.vector))
R+   res.wmF.II[i, 7:9] <- getCover(tmle_poolQg$EY_gstar1[["CIs"]], truth.wmF)
R+ }             
\end{CodeInput}
\end{CodeChunk}
	
\begin{table}[h]
	\begin{center}
		\caption{Simulation study 1. Simulation-based performance of TMLE, IPTW, Gcomp estimators with stochastic exposures over 200 repetitions of the simulation, when the working model holds ($\psi_0 = 55.57\%$) and when the working model is not a reasonable approximation ($\psi_0 = 55.78\%$). TMLE-Ia indicates both the outcome regression and the treatment mechanism are adjusted at the community-level. TMLE-Ib uses the individual-level outcome regression and the community-level treatment mechanism. TMLE-II indicates both are adjusted at the individual-level. IPTW-I and Gcomp-I indicate the use of community-level treatment and community-level outcome, respectively. IPTW-II and Gcomp-II indicate the use of individual-level treatment and individual-level outcome, respectively. For each estimator, the columns denote $\hat{\psi}$ as the average point estimate, "Bias" as the absolute difference between the estimate $\hat{\psi}$ and the truth $\psi$, $\hat{\sigma}$ as the average standard error estimate, rMSE as the root mean squared error, and "Cover" as the proportion of times that the truth falls within the 95\% CI. All outcome and treatment mechanisms are correctly specified. All reported bias, SE, rMSE and Coverage are multiplied by 100.}
		\begin{tabular}{@{}lllllllllllll@{}}
			\toprule \toprule
			& \multicolumn{5}{c}{\textbf{Working Model Holds}} & & \multicolumn{5}{c}{\textbf{Working Model Fails}} \\
			\cmidrule{2-6}  \cmidrule{8-12} 
			Estimator & $\hat{\psi}$ & Bias & $\hat{\sigma}$ & rMSE & Cover & & 
			$\hat{\psi}$ & Bias & $\hat{\sigma}$ & rMSE & Cover \\
			\midrule
			TMLE-Ia & 55.75 & 0.18 & 0.60 & 0.62 & 89.5 & & 
			56.47 & 0.69 & 1.36 & 1.52 & 86.5\\
			\midrule
			TMLE-Ib & 56.48 & 0.91 & 0.64 & 1.12 & 68.5 & & 
			56.50 & 0.72 & 1.60 & 1.75 & 89.5 \\
			\midrule
			TMLE-II & 55.71 & 0.13 & 0.39 & 0.41 & 84.0 & & 
			57.59 & 1.81 & 1.48 & 2.34 & 69.0 \\
			\midrule 
			IPTW-I  & 56.63 & 1.06 & 2.60 & 2.81 & 100 & &
			56.16 & 0.38 & 2.91 & 2.94 & 100 \\
			\midrule
			IPTW-II & 55.67 & 0.10 & 3.44 & 3.44 & 100 & & 
			57.23 & 1.45 & 4.23 & 4.47 & 100 \\
			\midrule
			Gcomp-I & 55.83 & 0.26 & 0.60 & 0.65 & 90.5 & & 
			56.44 & 0.67 & 1.36 & 1.51 & 85.5 \\
			\midrule
			Gcomp-II & 55.75 & 0.18 & 0.39 & 0.43 & 87.5 & &
			57.57 & 1.79 & 1.48 & 2.33 & 71.0 \\
			\bottomrule \bottomrule
		\end{tabular}
		\label{table:Simulation study 1}
	\end{center}
\end{table}

Results displayed in Table \ref{table:Simulation study 1} shows the comparison of the performance of the two TMLEs when the assumption of "no covariate interference" holds and the assumption fails badly. As predicted by theory, the community-level targeted estimator (TMLE-Ia), which is always based on the aggregated data, has a good performance in both situations with negligible bias. However, the coverage rates of its influence-curve-based confidence intervals are lower than nominal (89.5\% and 86.5\%) due to small variances. In this case, TMLE-Ib, which uses a pooled individual-level outcome regression and then a community-level stochastic intervention, performs slightly worse than TMLE-Ia. When the working model holds, we observe that the coverage rate of TMLE-Ib (68.5\%) has a notable decrease compared to that of the other estimators because of a relatively large bias. As expected, the individual-level targeted estimator (TMLE-II) is biased and its confidence interval coverage has an obvious decrease when the working model fails. Even though the working model is not a reasonable approximation, surprisingly, the IPTW using individual-level stochastic intervention (IPTW-II) provides a reasonable estimate.

\subsection[Simulation 2 - Static interventions]{Simulation 2 - Static interventions}\label{Simulation 2 - Static interventions}
We now consider another common simulation study with binary community-level exposure(s), which is commonly used in the study of HIV prevention and treatment. Similar to the previous simulation, we generate 200 samples of size $J = 100$ communities, each containing nj observation where $n_j \sim N (50, 10)$. 
The data generating mechanism is as follows.
\begin{gather*}
W_{1,n_j} \sim (Bern(0.6))_{i=1,...,n_j} \hspace{1cm} W_{2,n_j} \sim (N(0,1))_{i=1,...,n_j} \\
W_{1,n_j}^c = \frac{1}{n_j}\sum\limits_{i=1}^{n_j}{\bf W}_{1,n_j} \hspace{1cm} 
W_{2,n_j}^c = \frac{1}{n_j}\sum\limits_{i=1}^{n_j}{\bf W}_{2,n_j} \\
A_j \sim Bern(expit(W_{1,n_j}^c + 0.56W_{2,n_j}^c))
\end{gather*}

However, the mechanism differs in the outcome distribution:
\begin{itemize}
	\item \textbf{Case 1}: Working model is a reasonable approximation
	\[ Y_j \sim Bern(expit(0.15 + 0.3A_j + 0.1{\bf W}_{1,n_j}^c + 2{\bf W}_{1,n_j} + 0.9{\bf W}_{2,n_j})) \]
	
	\item \textbf{Case 2}: Working model is not a reasonable approximation
	\[ Y_j \sim Bern(expit(0.15 + 0.3A_j + 3{\bf W}_{1,n_j}^c - 0.9{\bf W}_{2,n_j}^c - 0.3{\bf W}_{1,n_j} + {\bf W}_{2,n_j})) \]
\end{itemize}

\begin{table}[h]
	\begin{center}
		\caption{Simulation study 2. Performance of TMLE, IPTW, Gcomp estimators with binary exposures over 200 repetitions of the simulation, when the working model approximately holds ($\psi_0 = 4.16\%$) and when the working model does not hold ($\psi_0 = 3.71\%$). All outcome and treatment mechanisms are correctly specified. All reported bias, SE, rMSE and Coverage are multiplied by 100.}
		\begin{tabular}{@{}lllllllllll@{}}
			\toprule \toprule
			& \multicolumn{4}{c}{\textbf{Working Model Holds}} & & \multicolumn{4}{c}{\textbf{Working Model Fails}} \\
			\cmidrule{2-5} \cmidrule{7-10} 
			Estimator & Bias & $\hat{\sigma}$ & rMSE & Cover & & 
			Bias & $\hat{\sigma}$ & rMSE & Cover \\
			\midrule
			TMLE-Ia & 0.03 & 1.15 & 1.16 & 95.0 & & 
			0.03 & 1.07 & 1.08 & 92.5 \\
			\midrule
			TMLE-Ib & 0.16 & 1.16 & 1.17 & 95.0 & & 
			0.25 & 1.24 & 1.26 & 97.5 \\
			\midrule
			TMLE-II & 0.01 & 1.14 & 1.14 & 95.00 & & 
			0.04 & 1.22 & 1.22 & 96.0 \\
			\midrule 
			IPTW-I & 0.02 & 3.79 & 3.79 & 100 & & 
			0.06 & 3.46 & 3.46 & 100 \\
			\midrule
			IPTW-II & 0.04 & 15.99 & 15.99 & 100 & & 
			0.02 & 17.56 & 17.56 & 100 \\
			\midrule
			Gcomp-I & 0.03 & 1.15 & 1.16 & 95.5 & & 
			0.03 & 1.07 & 1.08 & 91.5 \\
			\midrule
			Gcomp-II & 0.01 & 1.14 & 1.14 & 95.0 & &
			0.04 & 1.22 & 1.22 & 96.0 \\
			\bottomrule \bottomrule		
		\end{tabular}
		\label{table:Simulation study 2}
	\end{center}
\end{table}

As before, table \ref{table:Simulation study 2} summarizes the performance of the estimators under different outcome generating distributions. First, TMLE-Ia performs well in both situations with negligible biases and great confidence interval coverages. As expected, TMLE-Ib performs similarly to TMLE-Ia when the working model holds, and worse than TMLE-Ia when it fails, in terms of bias and variance. TMLE-II, on the other hand, performs very well when the working model provides a reasonable approximation, but exhibits slight increases in bias and variance (and so a more conservative confidence interval) when the working model fails. Theoratically, TMLE-II uses lower dimensional objects with size $N =  \sum_{j=1}^J{N_j}$ and so may improve the finite sample efficiency if the working model holds. However, when the working model does not hold, the misspecification of both the outcome and treatment regressions will cause biased estimate and efficiency loss. Besides, both IPTW-I (with the community-level g) and IPTW- II (with the individual-level $g$) have larger variances compared to other estimators, and so provides 100\% coverage rates. It could be explained that the IPTW estimator has relatively large variability, despite the large sample size. In other words, the range of the estimated values of IPTW can be wide and results in a large variance.

\subsection[Simulation 3 - Stochastic interventions (N = 1)]{Simulation 3 - Stochastic interventions (N = 1)}\label{Simulation 3 - Stochastic interventions (N = 1)}
In this simulation, we study the special case where each community has only one observation (i.e., $N=1$) and the intervention is stochastic. As described in section \ref{Special case where one observation per community}, it's similar to data with only community-level baseline covariates (i.e., treat $(E, W) = E$). The data-generating distribution is described as follows:
\begin{gather*}
E_1 \sim Bern(0.5) \hspace{1cm} E_2 \sim Bern(0.3) \\
E_3 \sim N(0, 0.25) \hspace{1cm} E_4 \sim Unif(0, 1) \\
A | E_1, E_2, E_3, E_4 \sim N(0.86E_1 + 0.41E_2 - 0.34E_3 + 0.93E_4, 1) \\
Y | A, E_1, E_2, E_3, E_4 \sim N(3.63 + 0.11A - 0.52E_1 - 0.36E_2 + 0.12E_3 - 0.13E_4, 1)
\end{gather*}

Given a shift value and a truncation bound, the intervened exposure $A^*$ is distributed as:
\begin{gather*}
A_j^* = 
\begin{cases} A_j  + shift & exp\{0.5 * shift * (A_j - \mu(E_1, E_2, E_3, E_4) - \frac{shift}{2}\} > truncbd \\ 
A_j, & o.w. \end{cases} \nonumber \\
\text{where}  \hspace{0.5cm} 
\mu(E_1, E_2, E_3, E_4) = 0.86E_1 + 0.41E_2 - 0.34E_3 + 0.93E_4
\nonumber
\end{gather*}

Given a shift of 2 and a truncation bound of 10, the marginal treatment effect of the individual-based intervention is $\psi_0 = 3.505$. In the next step, we will explore the estimation performance of the targeted estimators with different choices of binarization methods and the number of bins. Also, we are interested in the performance of the estimators under different model specifications, including correctly specified and misspecified models for the outcome regression and the density of the conditional treatment distribution. Again, code to generate the example dataset is attached in the supplementary material.

\begin{figure}[h]
	\centering
	\includegraphics[width=\textwidth, height=9cm]{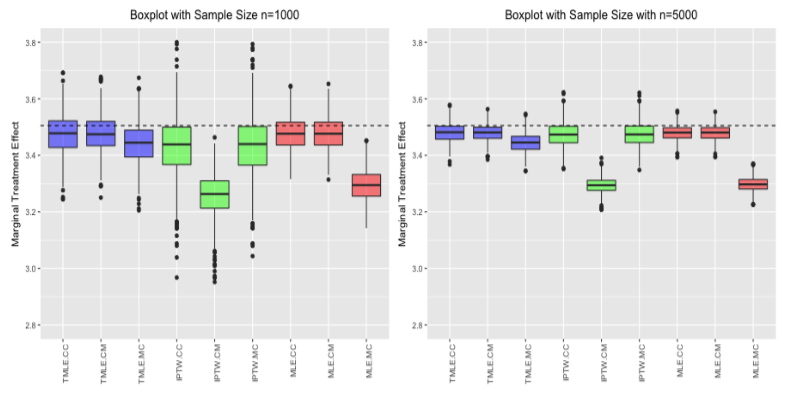}
	\caption{Box plots of the point estimates from three algorithms for sample sizes $n=1000$ (left) and $n=5000$ (right) in Simulation study 3. The x-axis denotes the combination of the estimator and the model specification. CC indicates correctly specified outcome regression and exposure mechanism. CM indicates the outcome regression is correctly specified, while the exposure mechanism misspecified. MC indicates the exposure mechanism is correctly specified, while the outcome regression misspecified. The dashed line indicates the true value $\psi_0 = 3.505$.}
	\label{figure:Simulation study 3}
\end{figure}

In figure \ref{figure:Simulation study 3}, the outcome $\bar{Q}_0(A,E)$ is estimated with the correctly specified main terms regression model and a misspecified regression model, only adjusting for $A$ and $E_3$. Besides, the stochastic exposure $g^*(a|E)$ is estimated with a correctly specified model, as well as a $g(a|E)$ misspecified model, only adjusting for $E_3$. The simulations results are consistent with the theoretical predictions. TMLE performs quite well if either the outcome regression or the exposure mechanism is correctly specified. IPTW exhibits low bias when $g^*(a|E)$ is correctly specified, but is biased otherwise. This bias decreases but does not disappear with an increase in sample size. Besides, IPTW has much higher variance than other estimators even with a correctly specified $g^*(a|E)$, which may be explained by practical positivity violations such as small $g^*(a|E)$ causes large weights on few individuals. Weight truncation could be a possible solution for this practical violation - We can implement more restrictive bounds on $g^*$ (in the simulation analysis, a less restrictive set of bounds of $[0.005, 1]$ is used). When the model for $\bar{Q}_0(A,E)$ is misspecified, MLE performs poorly in precision, but MLE is unbiased when $\bar{Q}_0(A,E)$ is correctly specified. Furthermore, sample size does help reduce variance. 

\section[Discussion]{Discussion}\label{Discussion}
The \pkg{tmleCommunity} package was developed to offer a flexible, easily customizable implementation of the TMLE algorithm for hierarchical data structure, along with community-level multivariate arbitrary interventions. The main class of causal parameters that is estimated by the package is the treatment specific mean effect, which can be easily extended to ATE. A neophyte only needs to supply the data and specify the data arguments \code{Ynode}, \code{Anodes}, \code{WEnodes} and \code{f_gstar1}. On that basis, experienced users can control the estimation procedure by providing the user-supplied regression models for $\bar{Q}_0$, $g_0$ and $g^*$, and choosing preferred methods allowed for arguments, such as the method dealing with hierarchical data, whether including hierarchical structure to estimate $\bar{Q}_0$, either linear or logistic fluctuation for targeting, and the TMLE targeting step method. Remarkably, \code{obs.wts} and \code{community.wts} can be used to correct for case-control sampling (when the outcome is rare). Besides, the \code{tmleCommunity} function can either internally estimate all factors of the likelihood, or use the values for $g_n$ and $g_n^*$  from the external estimation procedure through \code{h.g0_GenericModel} and \code{h.gstar_GenericModel}. The choices of data-adaptive machine learning techniques and other more advanced estimation methods can be specified in the \code{tmleCom_Options} function. 

Planned extensions to the package include several areas. First, we plan to include TMLE estimation of casual effects of multiple time-point interventions, adjusting for time-dependent covariates for hierarchical longitudinal data. Second, since considering only complete cases in the data is inefficient and may cause bias when missingness is informative. The package will then be extended to allow missingness on the outcome vector, so that the corresponding covariate information can be utilized for reducing bias and increasing efficiency in estimates. Third, as mentioned in section (\ref{Incorporating hierarchical structure for estimating outcome mechanism}), when one individual's outcome is affected by the individual-level covariates of the subset of other individuals from the same community, the strength of the "no covariate interference" assumption should be weakened by including this knowledge of dependence. Another planned addition to this package will so allow estimation of community-level TMLE under this setting. 

Additionally, this package estimates variances and standard errors through estimated influence curves. Double robustness makes these estimates asymptotically correct if both the outcome and treatment mechanisms are estimated consistently at reasonable rates, and conservative if only one of them is estimated consistently. However, variance estimation is difficult when violations or near violations of positivity happen in finite samples due to chance \citep{petersen_porter_gruber_wang_van_der_laan_2010}. This is usually a problem in small samples or when the exposure is continuous, since discretization of the support of the exposure could lead to lack of data in some bins. This sparsity results in poor finite sample performance, particularly for estimations of variances and confidence interval coverages, and even threatens valid inference. One alternative method for variance estimation is the non-parametric bootstrap, especially when central limit theorem may not apply due to sparsity. Thus, we plan to include the alternative variance estimates in the future. 

In this package, we use known stochastic interventions such as a shifted version of the current exposure mechanism $g_0$ given a known shift function. In practice, a stochastic intervention $g^*$ could also be unknown (i.e., not a function of $g_0$ anymore). If we consider the estimation of an optimal treatment rule where the rule is defined to maximize the mean outcome under the treatment, without cross validation, we will use the same information from the observed data to estimate both the user-specified mechanism $g^*$ and the mean outcome under the fitted mechanism, which may result in finite sample bias. According to \citet{van_der_laan_luedtke_2015} and \citet{luedtke_van_der_laan_2016}, the cross-validated TMLE (cv-TMLE) approach avoids empirical process conditions and for each sample split, it estimates an empirical mean over a validation sample, under a stochastic (or deterministic) intervention estimated based on the training sample. Therefore it may reduce finite sample bias, and including cv-TMLE in the package can be one of our future work.

\section[Answers to some frequently asked questions (FAQ)]{Answers to some frequently asked questions (FAQ)}\label{Answers to some frequently asked questions (FAQ)}

\textit{Can I call the \code{tmleCommunity} function a second time without having to re-do the estimation of exposure mechanism?}

Yes. Users can use command \code{result\$EY\_gstar1\$h.g0\_GenericModel} to obtain an object of \code{GenericModel} \pkg{R6} class containing the previously fitted models for ${P}(A|W, E)$ under observed mechanism $g_0$ (assuming the result of the first call to \code{tmleCommunity} is returned to the variable named \code{result}, and only one intervention function \code{f_gstar1} is user given). Similarly, an object \code{GenericModel} class containing the previously fitted models for ${P}(A|W, E)$ under intervention \code{f_gstar1} is returned as \code{result$EY_gstar1$h.gstar_GenericModel}. The two objects can be passed into the second call to \code{tmleCommunity} by specifying the values for \code{h.g0_GenericModel} and \code{h.gstar_GenericModel}, respectively. Assuming we are using the simulated data and the first estimation result in section \ref{Specification of estimation algorithms for treatment mechanisms}, the next code chunk illustrates how this is done.

\begin{CodeChunk}
\begin{CodeInput}
R> tmleCom_gc_default2 <- tmleCommunity(
R+   data = indSample.cA.cY, rndseed = 1, Ynode = "Y", Anodes = "A",
R+   WEnodes = c("W1", "W2", "W3", "W4"), Qform = "Y ~ W1 + W2 + A",
R+   h.g0_GenericModel = tmleCom_gc_default$EY_gstar1$h.g0_GenericModel,
R+   h.gstar_GenericModel = tmleCom_gc_default$EY_gstar1$h.gstar_GenericModel)
R> cbind(tmleCom_gc_default2$EY_gstar1$estimates,
R+       tmleCom_gc_default2$EY_gstar1$vars)
\end{CodeInput}
\end{CodeChunk}

\textit{Can I define and fit the multivariate conditional density under the user-specified intervention function directly without having to call the \code{tmleCommunity} function?}

Yes, the package provides an individual function named \code{fitGenericDensity} to define and fit regression models for the conditional density $\mathbb{P}(A = a|W = w)$ where $a$ is generated under a user-specified arbitrary (can be static, dynamic or stochastic) intervention function. Its arguments are similar to those for estimating treatment mechanisms in \code{tmleCommunity}, except hierarchical data structure is not supported in this function. Therefore, this function is purely for estimating the multivariate conditional density. 

With the same data set simulated in section \ref{Specification of estimation algorithms for treatment mechanisms}, we may be interested in the mean counterfactual outcome under a stochastic intervention $g^*$ where the observed $A$ is shifted to the left by the half of its mean. 
\begin{CodeChunk}
\begin{CodeInput}
R> define_f.gstar <- function(shift.rate, ...) {
R+   eval(shift.rate)
R+   f.gstar <- function(data, ...) {
R+     print(paste0("rate of shift: ", shift.rate))
R+     shifted.new.A <- data[, "A"] - mean(data[, "A"]) * shift.rate
R+     return(shifted.new.A)
R+   }
R+   return(f.gstar)
R+ }
R> f.gstar <- define_f.gstar(shift.rate = 0.5)
R> 
R> tmleCom_Options(maxNperBin = N, bin.method = "dhist", nbins = 8)
R> 
R> # Under current treatment mechanism g0
R> fit_gN <- fitGenericDensity(data = indSample.cA.cY, Anodes = "A",
R+                             Wnodes = c("W1", "W2", "W3", "W4"),
R+                             f_gstar = NULL, gform = gform.C)
R> # Under stochastic intervention gstar
R> fit_gstar <- fitGenericDensity(data = indSample.cA.cY, Anodes = "A",
R+                                Wnodes = c("W1", "W2", "W3", "W4"),
R+                                f_gstar = f.gstar, gform = gform.C)
\end{CodeInput}
\end{CodeChunk}

\textit{Are there any sample data provided in the package so that users can play analysis on them?}

Yes, the package comes with four sample datasets. \code{comSample.wmT.bA.bY\_list} is an example of a hierarchical data containing a community-level binary exposure with a Individual-Level binary outcome. And \code{indSample.iid.cA.cY\_list} is an example of a non-hierarchical data containing a continuous ex- posure with a continuous outcome. One non-hierarchical sample dataset is \code{indSample.iid.bA.bY.rareJ1\_list}, which contains a binary exposure with a rare outcome (i.e., independent case-control scenario where $J=1$). Beside, the data structure of another dataset \code{indSample.iid.bA.bY.rareJ2\_list} is identical to this of \code{indSample.iid.bA.bY.rareJ1\_list}, except that now the ratio of the number of controls to the number of case $J$ is 2.

\textit{Can the \pkg{tmleCommunity} package handle panel data transformation before performing TMLE analysis?}

Yes. The \code{panelData_Trans} function provides a wide variety of ways of data transformation for panel datasets, such as fixed effect and pooling model. It also allows users to only apply transformation on regressors of interests, instead of on the entire dataset. For example, before running the \code{tmleCommunity} function on the data set simulated in section \ref{Simulation 1 - Stochastic interventions} when the working model fails, we want to apply fixed effect transformation where the individual effect is introduced, then we can use
\begin{CodeChunk}
\begin{CodeInput}
R> pData.FE <- panelData_Trans(	
R>   data = comSample.wmF, xvar = c("E1", "E2", "W1", "W2", "W3", "A"),
R>   yvar = "Y", index = "id", effect = "individual",
R>   model = "within", transY = TRUE)
\end{CodeInput}
\end{CodeChunk}

Besides, we can keep the outcome variable fixed during the panel transformation by setting \code{transY = False}. Additional details can be found in the package manual  \url{https://github.com/chizhangucb/tmleCommunity/blob/master/tmleCommunity_Package_Documentation.pdf}. 

\section[Acknowledgments]{Acknowledgments}\label{Acknowledgments}

This work was supported by National Institutes of Health Director’s New Innovator Award Program DP2HD080350 (PI: Jennifer Ahern).
		
\bibliographystyle{jss}
\bibliography{tmleCommunity}

\end{document}